\documentclass[aps,prd,one-column,11pt,notitlepage]{revtex4-1}
\pdfoutput=1
\usepackage{setspace,braket}
\usepackage{amsmath, amssymb, bm, yfonts}
\usepackage{array,tabularx}
\usepackage{etoolbox}
\usepackage{graphicx}
    \makeatletter
    \patchcmd{\maketitle}{\@fpheader}{}{}{}
\makeatother
\usepackage{multirow}
\usepackage{longtable}
\DeclareMathOperator{\ve}{\varepsilon}

\def\ty{{\tilde y}}
\def\I{{\textrm I}}
\def\su{\mathfrak{su}}
\def\so{\mathfrak{so}}

\def \ts{\tilde{s}}

\def\be{\begin{equation}}
\def\ee{\end{equation}}

\def\Q{{\mathcal Q}}
\def\N{{\mathcal N}}

\def\W{{\mathcal W}}
\def\L{{\mathcal L}}

\def\L{{\mathcal L}}

\def\B{{\mathcal B}}
\def\A{{\mathcal A}}

\def\be{\begin{equation}}
\def\ee{\end{equation}}

\def\a{\alpha}
\def\b{\beta}

\def\d{\delta}
\def\e{\epsilon}

\def\l{\lambda}
\def\m{\mu}\def\n{\nu}\def\s{\sigma}\def\l{\lambda}

\def\bg{\bar{g}}

\def\beq{\begin{eqnarray}}\def\eeq{\end{eqnarray}}
\def\ba#1\ea{\begin{align}#1\end{align}}
\def\bg#1\eg{\begin{gather}#1\end{gather}}
\def\bm#1\em{\begin{multline}#1\end{multline}}
\def\bmd#1\emd{\begin{multlined}#1\end{multlined}}

\def\a{\alpha}
\def\b{\beta}
\def\c{\chi}

\def\d{\delta}

\def\e{\epsilon}
\def\ve{\varepsilon}

\def\l{\lambda}
\def\L{\Lambda}
\def\m{\mu}
\def\n{\nu}

\def\s{\sigma}
\def\t{\tau}

\def\nn{\nonumber}

\def\({\left(}
\def\){\right)}
\def\[{\left[}
\def\]{\right]}

\newcommand{\equ}[1]{Eq.~(\ref{#1})}
\newcommand{\sct}[1]{Sec.~(\ref{#1})}

\allowdisplaybreaks

\begin{document}

\title{Conformal embeddings and higher-spin bulk duals}
\author{Dushyant Kumar}
\email{dushyantkumar@hri.res.in}
\author{Menika Sharma}
\email{menikasharma@hri.res.in}
\affiliation{Harish-Chandra Research Institute, Chhatnag Road, Jhusi, Allahabad 211019, India}
\begin{abstract}{ It is well-known that conformal embeddings can be used to construct non-diagonal modular invariants for affine lie algebras. This idea can be extended to construct infinite series of non-diagonal modular invariants for coset CFTs.  In this paper, we systematically approach the problem of identifying higher-spin bulk duals for these kind of non-diagonal invariants. In particular, for a special value of the 't Hooft coupling, there exist a class of partition functions that have enhanced supersymmetry, which should be reflected in a bulk dual. As an illustration of this, we show that a partition function of an orthogonal group coset CFT has a $\N=1$ supersymmetric higher-spin bulk dual, in the 't Hooft limit. We also propose that two of the series of CFT partition functions, obtained from conformal embeddings, are equal, generalizing the well-known dual interpretation of the 3-state Potts model as a $\frac{SU(2)_3 \otimes SU(2)_1}{SU(2)_4}$ and also as a $\frac{SU(3)_1 \otimes SU(3)_1}{SU(3)_2}$ coset model. }
\end{abstract}
\maketitle


\section{Introduction}
The $W_N$ coset CFT and its relation with three-dimensional higher-spin Vasiliev theory \cite{Gaberdiel:2010} is a well-tested example of the AdS-CFT correspondence. This work has been extended in many directions. As of now, there is a plethora of coset CFTs with bulk duals. The duality has been shown to hold in the 't Hooft limit, as $N\rightarrow \infty$, as well as a semi-classical limit, where the central charge $c\rightarrow\infty$ \cite{Castro:2011}. Recent work on this duality addresses the embedding of the higher-spin Vasiliev theory into string theory and the structure of the unbroken symmetry algebra of string theory \cite{Gaberdiel:2014}. 

In the context of the duality between a Vasiliev higher-spin theory and a CFT, by and large, the duality maps a diagonal invariant of the CFT to the partition function of the bulk theory. However, a CFT often possesses a number of modular invariants. These come with varying spectra and one can, therefore, expect that their bulk duals should also be different. This is borne out by the few examples, that exist in the literature, of a duality between a non-diagonal CFT invariant and a higher-spin bulk theory~\cite{Gaberdiel:2014a,Beccaria,Creutzig:2014}. However, so far, there has been no systematic attempt to understand where different non-diagonal invariants of coset CFTs fit in the duality picture. We propose to address this question in this paper and Ref.~\cite{Sharma:2016}. 

For any coset CFT, it is a hard problem to classify all modular-invariant partition functions \cite{Beltaos:2010, Cappelli:1987}. 
Modular invariants of a coset CFT are intimately related to modular invariants of WZW models. Coset CFTs are of the form $G_s/H_x$ where the group $H$ is embedded in $G$, with $s$ and $x$ the levels of the affine groups $G$ and $H$ respectively.  The character of this coset model, also known as the branching function, is the coefficient of the expansion of the character of the affine group $G_s$ into the characters of the affine group $H_x$:
\be \label{branchDef}
	\c^G_{\m} = \sum_{\a} b_{\m\a} \c^H_{\a}.
\ee 
Here,  $b_{\m\a}$, the branching function of the coset CFT,  is a function of the modular parameter $\tau$ and carries an index $\m$ labelling the primary fields of the $G_s$  WZW model and an index $\a$ labelling the primaries of the $H_x$ WZW model.
The problem of classifying modular invariants of a coset CFT is, therefore, a problem of classifying the modular invariants of WZW models. However, although a complete classification of the modular invariants of WZW models remains elusive, it is easy to identify distinct classes of modular invariants. All WZW models have a diagonal invariant of the form
\be
	\sum_{\a} \left |\c_{\a}\right |^2,
\ee
referred to in the literature as an ``$A$''-type invariant. There is an important class of invariants called ``$D$''-type invariants that can be constructed from the diagonal partition function by modding out by a discrete symmetry of the WZW model. We discuss them in more detail in Ref.~\cite{Sharma:2016}. The third class consists of exceptional invariants or ``$E$''-type, constructed using a variety of methods. This class of invariants is the one that is difficult to classify.  However, there is a sub-class of $E$-type invariants that have a straightforward origin. These are the $E$-type invariants that result from conformal embeddings. 

An embedding $H_x \subset G_r$ is conformal when the central charges associated with the WZW models with gauge groups $H_x$ and $G_r$ are equal. If $H_x$ is conformally embedded in $G_r$ it implies that a character of $G_r$ can be expanded in terms of the characters of $H_x$ with constant coefficients. This means that in an expansion as in \equ{branchDef}, the $b_{\m\a}$ will be independent of $\tau$. As a result of this relation between their characters, a partition function of $G_r$ will result in a partition function of $H_x$. This idea has been widely utilised \cite{Walton:1988} to construct exceptional partition functions of many WZW models. 

There is a well-known procedure to construct modular invariants of a coset CFT once the modular invariants of WZW models are known. We review this procedure in Sec.~(\ref{EnumerateMIs}). Our aim is to study examples of modular invariants of coset CFTs that can be constructed because a constituent affine group of the coset is conformally embedded in a larger group. All conformal embeddings are known. Some conformal embeddings only appear for specific values of the ranks of the groups $G$ and $H$, however, others appear for generic values of the rank and specific values of the levels in terms of the rank. We are interested in the second class of embeddings because they result in infinite series of partition functions. Getting a series of invariants is particularly useful from the point of view of the bulk dual because it allows one to take a 't Hooft limit. However, since a conformal embedding always fixes the value of the level $(k)$  in terms of the rank $(N)$ of the gauge group $H$, the 't Hooft coupling given by $\lambda=N/(N+k)$ is fixed at a particular value. In all examples that we look at in this paper $\lambda$ is fixed at $1/2$ in the $N,k\rightarrow \infty$ limit.

It also turns out that partition functions generated in this way for different coset series, are not always different. Some of these coset series have the same central charges and by using the method of ``T-equivalence" \cite{Bowcock, Altschuler} one can find a relation between their branching functions. In Sec.~(\ref{SeriesEquivalence}) we give examples of such equivalences. 

For identifying the bulk dual of these cosets, we need to find the symmetry algebra of these partition functions. In general, the symmetry algebra of a non-diagonal invariant of a coset model is either the same as the symmetry algebra of the diagonal invariant or an extension thereof. We show in Sec.~(\ref{EnumerateMIs}) that some of the partition functions constructed using the conformal embedding technique can be interpreted as diagonal partition functions of a different coset model. These partition functions, as a consequence, have a larger symmetry algebra than the original partition functions. In Sec.~(\ref{BulkDuals}), we discuss a special case of this phenomena, for the coset model
\be
	\frac{SO(2N)_{2N-2} \otimes SO(2N)_1}{SO(2N)_{2N-1}}\,.
\ee
Cosets with orthogonal gauge groups and their symmetry algebras have been discussed in the literature before \cite{Gaberdiel:2011nt, Candu, Candu:2012, Creutzig:2012,Ferreira:2014,Ahn:2011}, but this particular case is new. Its importance is related to the fact that taking a non-diagonal invariant of this coset model, the symmetry algebra is boosted from ${\cal N}=0$ to ${\cal N}=1$ supersymmetry. This mirrors what happens for the $\frac{SU(N)_{N} \otimes SU(N)_1}{SU(N)_{N+1}}$ coset \cite{Beccaria} and is part of a more general phenomena of enlarged supersymmetry at $\lambda=1/2$ for particular non-diagonal invariants of coset models.

This paper is organized as follows. In \sct{EnumerateMIs}, we construct exceptional-type non-diagonal modular invariants of coset CFTs, and identify the type of bulk duals they can have based on their symmetry algebras. In \sct{SeriesEquivalence}, we provide evidence that two {\it a priori} distinct series of modular invariants are actually equal. Then in \sct{BulkDuals} we study a particular non-diagonal invariant and show that it has a bulk dual with a higher-spin symmetry algebra consisting of supermultiplets with spin $(2k+\frac{3}{2}, 2k+2)$, where $k=0,1,2,\cdots$. As mentioned before, this is an analog of the duality between a non-diagonal invariant of the coset $\frac{SU(N)_{N} \otimes SU(N)_1}{SU(N)_{N+1}}$ and a higher spin bulk with symmetry algebra consisting of fields of spin $(k+\frac{3}{2}, k+2)$ --- however, the details of these dualities are different.  In \sct{discussion} we summarise our work and list some open problems related to it.

\section{Modular invariants of the coset CFT from embeddings \label{EnumerateMIs}}
In this section, we construct non-diagonal modular invariants of cosets of the form
\be 
\label{genSeries}
	\frac{H_x \otimes H_y}{H_{x+y}}
\ee
where $H$ denotes one of the classical Lie groups. The ``minimal model'' coset 
\be 
\label{coset}
\frac{SU(N)_k \otimes SU(N)_1}{SU(N)_{k+1}}
\ee
is the most studied example of this class of cosets.
We will show in \sct{supercosets}, that a particular type of non-diagonal invariant of the coset in \equ{genSeries} can be reinterpreted as diagonal invariant of a different coset model and this interpretation determines its symmetry algebras, which we discuss in \sct{exs}.

Modular invariants for coset models of the kind in \equ{genSeries} are built up from modular invariants of the affine group $H$. A general partition function of the coset CFT, for which the level $y$ is fixed at one, can be written as
\be \label{genPart}
	{\cal Z} =\frac{1}{\ell} \sum_{\m,\n} b_{\m\n} { \cal M}_{(\m,\m^\prime);(\n,\n^\prime)} \overline b_{\m^\prime\n^\prime} \,.
\ee
Here, $\ell$ denotes the order of the outer automorphism group of the affine algebra of $H$. The matrix ${ \cal M}$ is given by
\be
{ \cal M}_{(\m,\m^\prime);(\n,\n^\prime)} = {\mathit m}^I_{\m;\m^\prime }{ \mathit m}^{II}_{\n;\n^\prime},
\ee
such that $\sum_{\m}\chi_\m { \mathit m}^I_{\m;\m^\prime } \overline{\chi}_{\m^\prime}$ and $\sum_{\n}\chi_\n { \mathit m}^{II}_{\n;\n^\prime } \overline \chi_{\n^\prime}$ are partition functions of the WZW models $H_x$ and $H_{x+y}$ respectively. For this paper, we are interested in those partition functions for which at least one of ${ \mathit m}^{I}$ or ${ \mathit m}^{II}$ is not equal to the identity matrix so that the resultant partition function in \equ{genPart} is non-diagonal. 

As stated in the introduction, it is customary to refer to the diagonal partition function of a WZW model as an $A$-type model, while the non-diagonal partition functions are referred to as $D$ or $E$ type models depending on their method of construction \cite{Cappelli:1987}. For the coset partition function, the same notation is adapted. Thus, the diagonal partition function of a coset , for which the matrices ${ \mathit m}^{I}$ and ${ \mathit m}^{II}$ are both equal to the identity matrix, is denoted as the $AA$ partition function. A partition function for which ${ \mathit m}^{I}$ is the diagonal matrix but ${ \mathit m}^{II}$ is a non-diagonal matrix corresponding to an $E$-type invariant of a WZW model, is referred to as an $AE$ partition function.

\begin{table}[!t]
 \begin{center}
\renewcommand{\arraystretch}{1.5}
 \begin{ruledtabular}
 \begin{tabular}{  c c l c c c l}
 \multirow{2}{*}{$SU(N)_N$} &  \multirow{2}{*}{$\subset$} &  \multirow{2}{*}{$SO(N^2-1)_1$} &~~& $SO(N)_{N-2}$ & $\subset$ & $SO(\frac{N(N-1)}{2})_1$ \\
\cline{5-7}
 & & &~~&$Sp(2N)_{N+1}$ &$\subset$& $SO(\frac{N(2N+1)}{2})_1$ \\
 \hline
 $SU(N)_{N+2}$ &$\subset$&  $SU(\frac{N(N+1)}{2})_1$ &~~&  $SO(N)_{N+2}$ &$\subset$& $SO(\frac{(N-1)(N+2)}{2})_1$ \\
 \hline
  $SU(N)_{N-2}$ &$\subset$&  $SU(\frac{N(N-1)}{2})_1$ &~~&  $Sp(2N)_{N-1}$ &$\subset$& $SO(\frac{(N-1)(2N+1)}{2})_1$ \\
 \end{tabular}
 \end{ruledtabular}
 \caption{Conformal embeddings that are present at generic values of the rank for the classical Lie groups.}
 \label{t:1}
  \end{center}
 \end{table}

We will construct modular invariants for the coset model that result from conformal embeddings. For this, we first need to know the conformal embeddings for WZW models. In Table~(\ref{t:1}), we list all conformal embeddings for the groups $SU(N)$,  $SO(N)$ and $Sp(2n)$ that are present for generic values of the group rank. Each embedding in Table~(\ref{t:1}) gives rise to partition functions for two different coset models: one for which $H_x$ can be chosen to be the embedded group and the second for which the $H_{x+y}$ can be chosen to be the embedded group. For example, the embeddings in row two and three of Table~(\ref{t:1}) for the $SU(N)$ group give rise to non-diagonal partition functions of the coset models:
\beq \label{SUSeries}
	 {\rm Series~I:}&~~ \frac{SU(N)_{N+1} \otimes SU(N)_1}{SU(N)_{N+2}},\nn\\
	 {\rm Series~II:}&~~\frac{SU(N+1)_{N-1} \otimes SU(N+1)_1}{SU(N+1)_{N}},\nn \\
	 {\rm Series~III:} &~~\frac{SU(N)_{N+2} \otimes SU(N)_1}{SU(N)_{N+3}},\nn \\
	 {\rm Series~IV:}&~~\frac{SU(N+1)_{N-2} \otimes SU(N+1)_1}{SU(N+1)_{N-1}}.
\eeq
Not all the partition functions obtained in this way are distinct. The central charges for Series~I and Series~II are the same, implying that there may exist a relation between the branching functions of these cosets. In fact, we present evidence in Section~(\ref{SeriesEquivalence}) that the partition functions obtained via conformal embeddings for these two series are the same. This kind of equivalence happens for other series of coset models as well. 
The following two series
\beq \label{SOSeries}
	{\rm Series~V:} &~~\frac{SO(N)_{N-2} \otimes SO(N)_1}{SO(N)_{N-1}},\nn \\
	 {\rm Series~VI:}&~~ \frac{SO(N-1)_{N} \otimes SO(N-1)_1}{SO(N-1)_{N+1}}.
\eeq
also have identical central charges.

\subsection{Interpretation as a diagonal invariant \label{supercosets}}
Of particular interest are non-diagonal partition functions of \equ{genSeries} that are equal to diagonal partition functions of a different coset model. This is possible if it is the group $H_x$ that is conformally embedded in another group $G_r$ . In this case, a specific partition function of \equ{genSeries} is equal to the diagonal $(AA)$ partition function of the coset model:
\be \label{genSeriesSu}
	\frac{G_r \otimes H_y}{H_{x+y}}\,.
\ee
This comes about in the following way. A conformal embedding implies that the characters of $G_r$ can be expanded in the characters of $H_x$ as
\be \label{charRelation}
	\chi^{G_r}_\xi = c_{\xi \m} \chi^{H_x}_\m\,,
\ee
where the 	$c_{\xi \m}$ are constants independent of $\tau$. The characters are, in general, functions of $\tau$ and $z$ where $z$ is in the Cartan subgroup of the associated group. Therefore, the diagonal partition function of $G_r$ 
\be
	{\mathcal Z}^G_{\textrm {diagonal}} = \sum_\xi \chi^{G_r}_\xi \bar{ \chi}^{G_r}_\xi 
\ee
results in a non-diagonal partition function of $H_{x}$:
\be
	{\mathcal Z}^H_{\textrm {non-diagonal}} = \sum_\xi \left |c_{\xi \m} \chi^{H_x}_\m \right|^2 \,.
\ee
The branching function $B_{\xi \n {\theta}}$ for the coset model in \equ{genSeriesSu} obeys the following equation
\be
\label{BGH}
	\chi^{G_r}_{\xi} \chi^{H_y}_{\theta} = \sum_{\n} B_{\xi \n {\theta}} \chi_\n^{H_{x+y}}\,.
\ee
Similarly, the branching function $b_{\m \n \theta}$ of the coset model in \equ{genSeries} obeys
\be
	\chi^{H_x}_{\m} \chi^{H_y}_{\theta} = \sum_{\n} b_{\m \n \theta} \chi_\n^{H_{x+y}}\,.
\ee
Summing on both sides over the index $\m$ after multiplying each side with the coefficients  $c_{\xi \m}$ gives
\be
	\sum_{\m} c_{\xi \m}  \chi^{H_x}_{\m} \chi^{H_y}_{\theta} = \sum_{\m} \sum_{\n} c_{\xi \m}  b_{\m \n \theta} \chi_\n^{H_{x+y}}\,.
\ee
Using the relations in Eqs.~(\ref{charRelation}) and (\ref{BGH}) and the linear independence of the characters $\chi_\n^{H_{x+y}}$ results in
\be \label{branRelations}
	B_{\xi \n {\theta}}  = \sum_{\m} c_{\xi \m}  b_{\m \n \theta} 
\ee
As for the WZW models, the above relation implies that the $AA$ diagonal partition function of the coset model in \equ{genSeriesSu} yields a non-diagonal partition function of the coset model in \equ{genSeries}.

\begin{table}[b]
 \begin{center}
 \renewcommand{\arraystretch}{1.5}
 \begin{ruledtabular}
 \begin{tabular}{  c c l c c c l}
 \multirow{2}{*}{$SU(N)_N$} &  \multirow{2}{*}{$\subset$} &  \multirow{2}{*}{$SO(N^2-1)_1$} &~~& $SO(N)_{N-2}$ & $\subset$ & $SO(\frac{N(N-1)}{2})_1$ \\
\cline{5-7}
 & & &~~&$Sp(2N)_{N+1}$ &$\subset$& $SO(\frac{N(2N+1)}{2})_1$ \\
  \end{tabular}
  \end{ruledtabular}
 \caption{Conformal embeddings that give rise to supersymmetric models}
 \label{t:2}
  \end{center}
 \end{table}

\subsection{Extended symmetry algebra \label{exs}}

The currents of the coset model are those fields of the $G_r \otimes H_y$ WZW model that commute with all fields of $H_{x+y}$. To find the extended symmetries of the coset model in \equ{genSeriesSu}, we use the method in Ref.~\cite{Bowcock}. 
Although, this method does not give all the currents associated with the coset model, it allows us to quickly identify the type of symmetry algebra associated with a model. For example, it will allow us to determine whether the algebra is supersymmetric.

Following the usual convention, we denote the algebra associated with a affine group with small case letters, so {\it e.g} $\hat h_x$ denotes the algebra associated with the group $H_x$. The goal is to identify a field $\phi$ in the representations of $\hat g_r$ that commute with all fields of $\hat h_y$. Such a field is denoted as a ``h-scalar'' and can be shown to extend the conformal algebra. To identify such h-scalars, it is convenient to consider cases according to the integral part of the conformal dimension $\Delta_\phi$ of the field $\phi$. If the conformal dimension of the field is between $0$ and $1$, then $\phi$ can just be the vacuum state. For $\Delta_\phi$ to lie between $1$ and $2$, it is a necessary condition that a field transforming in the adjoint representation of $\hat h$ exist in the decomposition of representations associated with $\hat g_r$. If such a field exists, it has a corresponding h-scalar whose conformal weight is given by
\beq
	\Delta_\phi = 1 + \frac{Q_\psi}{Q_\psi + 2 y}
\eeq
where $\Q_\psi$ is the quadratic Casimir of the adjoint representation of $H$.
Using the above equation we can find the dimension of the extra symmetry current for partition functions associated with the conformal embeddings listed in Table~(\ref{t:1}). As mentioned above, for this symmetry current to be present, one also needs to check that the adjoint representation of $\hat h$ appears in a decomposition of the allowed representations of $g_r$. In Table~(\ref{t:2}), we list the conformal embeddings that result in partition functions having a current of spin $3/2$. 
All the groups listed in Table~(\ref{t:2}) are embedded in a group of the form $SO(M)_1$, whose WZW model can be realised in terms of free fermions. 

Not all embeddings in Table~(\ref{t:1}) give rise to partition functions with an extra symmetry current with conformal dimension in the range $1$ to $2$. While some of these models have parafermionic currents at special values of the rank $N$, these currents are not generic to the whole series. Indeed, some of these partition functions may just have the symmetry algebra of the original theory, without any extension.


\section{Equivalent partition functions \label{SeriesEquivalence}}
The central charges associated with Series~I and Series~II are equal and given by 
\begin{equation}
c_N =  \frac{(N-1)(3N+2)}{4N+2}.
\end{equation}
However, in general these two series are not isomorphic. The branching functions associated with these series, and hence {\it e.g} the diagonal partition functions, are known to be different. Our claim in this section is that a particular $AE$ partition function of Series~$\I$ corresponding to the conformal embedding $SU(N)_{N+2}\subset SU(N(N+1)/2)_1$ is equal to a $EA$ type partition function of Series~$\I\I$ coming from conformal embedding $SU(N+1)_{N-1}\subset SU(N(N+1)/2)_1$. We also propose that a similar equivalence of partition functions takes place for Series~V and Series~VI. For much of this paper, we will concentrate on Series~VI, however, from the point of view of equivalence of partition functions, Series~I and Series~II present the more interesting case. Therefore, in this section we will focus on these series.

The equivalence is well-known for the first members of the two series, {\it i.e} for the $N=2$ case. The partition function of coset $\tfrac{SU(2)_3\otimes SU(2)_1}{SU(2)_4}$ corresponding to the embedding $SU(2)_4\subset SU(3)_1$ is  
\be
\mathcal{Z}^{\I}_{AD} = \mathcal{Z}^{\I}_{AE} =  \, \big | \, b_{0}^{\I}+b_{3}^{\I} \, \big | ^{2}\,+\, \big | \, b_{\frac{2}{5}}^{\I}+b_{\frac{7}{5}}^{\I} \, \big |^{2}\,+\,2\big | \,b_{\frac{2}{3}}^{\I} \, \big |^{2}\,+\,2 \big | \,b_{\frac{1}{15}}^{\I}\, \big |^{2}\,.
\label{eq:AEsu2}
\ee
Here and subsequently, the partition function depends on $q=\exp(2 \pi i \t)$. The superscript $\I$ on the branching function $b$ indicates that it is a branching function of Series~$\I$. For ease of notation, the branching functions in this case are labelled by the conformal weight of the coset primary field, rather than by the weights of the primaries of the constituent WZW models. 
The partition function of the coset $\tfrac{SU(3)_1\otimes SU(3)_1}{SU(3)_2}$ corresponding to the trivial embedding $SU(3)_1\subset SU(3)_1$ is simply the diagonal partition function 
\be
\mathcal{Z}^{\I \I}_{AA}  = \mathcal{Z}^{\I \I}_{EA} =  {\big |\,b_{0}^{\I\I}\, \big |}^{2}\,+\, {\big |\,b_{\frac{2}{5}}^{\I\I} \,\big |}^{2}\,+\,2 {\big|\,b_{\frac{2}{3}}^{\I\I\,}\big |}^{2}\,+\,2  {\big | \,b_{\frac{1}{15}}^{\I\I} \, \big| }^{2}\,.
\ee
It can be shown that the branching functions of the two coset models are related as
\begin{alignat}{2}
b_{0}^{\I\I} & = b_{0}^{\I}+b_{3}^{\I} \,, &\quad \quad
b_{\frac{2}{3}}^{\I\I} &=  b_{\frac{2}{3}}^{\I} \,,
\nonumber\\
b_{\frac{2}{5}}^{\I\I} & = b_{\frac{2}{5}}^{\I}+b_{\frac{7}{5}}^{\I}  \,,&\quad \quad
b_{\frac{1}{15}}^{\I\I} & =  b_{\frac{1}{15}}^{\I} \,.
\end{alignat}
and, therefore, that  $\mathcal{Z}^\I_{AE}=\mathcal{Z}^{\I\I}_{EA}$. 
In this section, we propose that this equivalence extends to the whole series. Although, we will not provide a complete proof of this, we provide several pieces of evidence to support this proposal. First, we show in \sct{N3} that this equivalence holds for $N=3$, by explicitly calculating the branching functions. Then in \sct{sec:tequi} we show that these coset series are dual or $T$-equivalent \cite{Bowcock, Altschuler}. $T$-equivalence implies that there exist some relations between the branching functions of the equivalent coset models and that, therefore, a partition function of one can be the same as a partition function of another. In \sct{prop} we provide an exact statement of the equivalence.

\subsection{The $N=3$ case \label{N3}}

The $AE$ partition function of the coset $\frac{SU(3)_4\otimes SU(3)_1}{SU(3)_5}$ corresponding to the conformal embedding $SU(3)_5\subset SU(6)_1$ can be constructed using the general form in \equ{genPart} and the $E$-type partition function for the $SU(3)_5$ partition function as given, for example, in \cite{Walton:1988}. The exact form of the partition function is
\begin{alignat}{1}
\mathcal{Z}^{\I}_{AE}~=~ \big |&\, b_{0}^{ \I}+b_{3}^{\I}\,\big|^{2}\,+\,2\big |\,b_{\frac{3}{28}}^{\I}+b_{\frac{87}{28}}^{\I}\,\big|^{2}+2\,\big |\,b_{\frac{5}{84}}^{\I}+b_{\frac{173}{84}}^{\I}\,\big|^{2} + 2\,\big |\,b_{\frac{2}{21}}^{\I}+ b_{\frac{65}{21}}^{\I}\,\big|^{2}+\big |\,2\,b_{\frac{39}{28}}^{\I}\,\big|^{2}
\nonumber \\
&~+~\big |\,b_{\frac{1}{7}}^{\I}+b_{\frac{36}{7}}^{\I}\,\big|^{2}+2\big |\,b_{\frac{29}{84}}^{\I}+b_{\frac{113}{84}}^{\I}\,\big|^{2} +  \big |\,b_{\frac{3}{7}}^{\I}+b_{\frac{10}{7}}^{\I}\,\big|^{2}+2\big |\,b_{\frac{11}{21},+}^{\I}+
b_{\frac{32}{21}}^{\I}\,\big|^{2}+\big |\,2\,b_{\frac{9}{4}}^{\I}\,\big|^{2}\nonumber \\
&~+~2\big |\,b_{\frac{11}{21},-}^{\I}+b_{\frac{116}{21}}^{\I}\,\big|^{2} +  \big |\, 2\,b_{\frac{19}{28}}^{\I}\,\big|^{2} +2\big |\,b_{\frac{65}{84},+}^{\I}+b_{\frac{149}{84}}^{\I}\,\big|^{2}+2\big |\,b_{\frac{65}{84},-}^{\I}+b_{\frac{233}{84}}^{\I}\,\big|^{2}\nonumber \\
 &~+~  2\big |\,b_{\frac{17}{21}}^{\I}+b_{\frac{38}{21}}^{\I}\,\big|^{2}+2\big |\,b_{\frac{6}{7}}^{\I}+b_{\frac{27}{7}}^{\I}\,\big|^{2}+2\big |\,b_{\frac{11}{12}}^{\I}+b_{\frac{59}{12}}^{\I}\,\big|^{2} +  2\big |\,b_{\frac{5}{3}}^{\I}+b_{\frac{20}{3}}^{\I}\,\big|^{2}\,.
 \label{eq:AE5}
\end{alignat}
Here, again, we have labelled the fields solely by their coset conformal dimension, except where we need to distinguish the fields for the identities listed later in this section, in which case we have included the $W_3$ charge denoted by $\pm$. The corresponding $SU(3)_4$ and $SU(3)_5$ weights for the fields present in the partition function appear in Table~(\ref{apptab1}) in Appendix~(\ref{seriesEx}). 

The $EA$ partition function of the coset $\frac{SU(4)_2 \otimes SU(4)_1}{SU(4)_3}$ corresponding to the embedding $SU(4)_2\subset SU(6)_1$ is
\begin{alignat}{1}
\label{eq:EA5}
\mathcal{Z}^{\I\I}_{EA} ~ = ~ \big |&\, b_{0}^{\I\I}+b_{3}^{\I\I}\,\big |^{2}+2\big |\,b_{\frac{3}{28}}^{\I\I}+b_{\frac{59}{28}}^{\I\I}\,\big |^{2}+2\big |\,b_{\frac{5}{84}}^{\I\I}\,\big |^{2}+\big |\,2 \,b_{\frac{9}{4}}^{\I\I}\,\big |^{2}+2\big |\,b_{\frac{2}{21}}^{\I\I}\,\big |^{2}+\big |\,b_{\frac{1}{7}}^{\I\I}+b_{\frac{8}{7}}^{\I\I}\,\big |^{2}\nonumber\\
 &~+~ 2\big |\,b_{\frac{29}{84}}^{\I\I}\,\big |^{2}+\big |\,b_{\frac{3}{7}}^{\I\I}+b_{\frac{10}{7}}^{\I\I}\,\big |^{2}+4\big |\,b_{\frac{11}{21}}^{\I\I}\,\big |^{2}+\big |\,2\, b_{\frac{19}{28}}^{\I\I}\,\big |^{2}+4\big |\,b_{\frac{65}{84}}^{\I\I}\,\big |^{2}+2\big |\,b_{\frac{17}{21}}^{\I\I}\,\big |^{2}\nonumber\\
 &~+~ 2\big |\,b_{\frac{6}{7}}^{\I\I}+b_{\frac{13}{7}}^{\I\I}\,\big |^{2}+2\big |\,b_{\frac{11}{12}}^{\I\I}\,\big |^{2}+\big |\,2\,b_{\frac{39}{28}}^{\I\I}\,\big |^{2}+2\big |\,b_{\frac{5}{3}}^{\I\I}\,\big |^{2}\,.
\end{alignat}
The corresponding $SU(4)_2$ and $SU(4)_3$ weights for the fields in the partition function appear in Table~(\ref{apptab2}). 

The branching functions of the two coset models are related by the following identities:
\begin{equation}
\label{n3Identities}
\begin{aligned}[c]
b_{0}^{\I\I}+b_{3}^{\I\I}&~=~b_{0}^{\I}+b_{3}^{\I} \,, \\
b_{\frac{3}{28}}^{\I\I}+b_{\frac{59}{28}}^{\I\I}&~=~b_{\frac{3}{28}}^{\I}+b_{\frac{87}{28}}^{\I} \,,\\
b_{\frac{1}{7}}^{\I\I}+b_{\frac{8}{7}}^{\I\I}&~=~b_{\frac{1}{7}}^{\I}+b_{\frac{36}{7}}^{\I}  \,,\\
b_{\frac{3}{7}}^{\I\I}+b_{\frac{10}{7}}^{\I\I}&~=~b_{\frac{3}{7}}^{\I}+b_{\frac{10}{7}}^{\I} \,, \\
b_{\frac{6}{7}}^{\I\I}+b_{\frac{13}{7}}^{\I\I}&~=~b_{\frac{6}{7}}^{\I}+b_{\frac{27}{7}}^{\I}\,,  \\
b_{\frac{9}{4}}^{\I\I} &~=~b_{\frac{9}{4}}^{\I}  \,,
\end{aligned}
\qquad
\begin{aligned}[c]
b_{\frac{2}{21}}^{\I\I}&~=~b_{\frac{2}{21}}^{\I}+b_{\frac{65}{21}}^{\I}\,, \\
b_{\frac{29}{84}}^{\I\I}&~=~b_{\frac{29}{84}}^{\I}+b_{\frac{113}{84}}^{\I}\,, \\
b_{\frac{11}{21}}^{\I\I}&~=~b_{\frac{11}{21},-}^{\I}+b_{\frac{116}{21}}^{\I}  \,,\\
b_{\frac{17}{21}}^{\I\I}&~=~b_{\frac{17}{21}}^{\I}+b_{\frac{38}{21}}^{\I}\,, \\
b_{\frac{11}{12}}^{\I\I}&~=~b_{\frac{11}{12}}^{\I}+b_{\frac{59}{12}}^{\I} \,,\\
b_{\frac{5}{3}}^{\I\I}&~=~b_{\frac{5}{3}}^{\I}+b_{\frac{20}{3}}^{\I} \,,\\
\end{aligned}
\qquad
\begin{aligned}[c]
b_{\frac{65}{84}}^{\I\I}&~=~b_{\frac{65}{84}, +}^{\I}+b_{\frac{149}{84}}^{\I} \,,  \\
b_{\frac{65}{84}}^{\I\I}&~=~b_{\frac{65}{84},-}^{\I}+b_{\frac{233}{84}}^{\I}  \,,\\
b_{\frac{5}{84}}^{\I\I}&~=~b_{\frac{5}{84}}^{\I}+b_{\frac{173}{84}}^{\I} \,,  \\
b_{\frac{11}{21}}^{\I\I}&~=~b_{\frac{11}{21}, +}^{\I}+b_{\frac{32}{21}}^{\I} \,, \\
b_{\frac{19}{28}}^{\I\I}&~=~b_{\frac{19}{28}}^{\I} \,, \\
b_{\frac{39}{28}}^{\I\I} &~=~b_{\frac{39}{28}}^{\I}   \,.
\end{aligned}
\end{equation}
We have derived these identities by computing the branching functions for both coset models. The $q$-series expansions for these branching functions are given in Tables~(\ref{apptab1}) and (\ref{apptab2}).
Using these identities, we conclude  that $\mathcal{Z}^\I_{AE}=\mathcal{Z}^{\I\I}_{EA}$. 
Note that identities of this type can only work if the minimum conformal dimension of fields associated with branching functions appearing in the L.H.S is the same as the minimum conformal dimension of field associated with branching functions appearing in the R.H.S. Further, all fields appearing in a identity should have conformal dimensions differing by integers. This observation makes it easy to predict what identities can hold for the $N>3$ cases. We have numerically checked the equivalence of partition functions of the two coset series up till the $N=5$ case.
\subsection{$T$-equivalence}
\label{sec:tequi}
In this section, we show that the Series~I and Series~II cosets constitute a dual pair, following the method in Refs.~\cite{Bowcock, Altschuler}. 
We use lower-case letters to denote the affine algebra associated with a WZW group, for example, $\hat{g}$ denotes the affine algebra for the group $G_x$.
Then, the two cosets $\hat{g}/\hat{h}$ and $\hat{g}'/\hat{h}'$ are said to be $T$-equivalent
or dual if there exists an affine algebra $\hat{g}_e$ such that $\hat{g}\oplus\hat{h}^{'}$
and $\hat{g}^{'}\oplus\hat{h}$ are both conformally embedded in $\hat{g}_e$. Because of the conformal embedding the stress-energy tensors of these algebras are related as 
\be
	T_{\hat{g}_e} = T_{\hat{g}} + T_{\hat{h}'} = T_{\hat{g}'} + T_{\hat{h}}\,.
\ee
From this it follows that $T_{\hat{g}/\hat{h}}= T_{\hat{g}'/\hat{h}'}$. $T$-equivalent cosets, therefore, have equivalent stress-energy tensors.

As an example, let us first show that Series~V and Series~VI of \equ{SOSeries} are $T$-equivalent. For these two series we have 
\begin{equation}
\begin{aligned}[c]
\hat{g} & = \so(N)_{N-2} \oplus \so(N)_1\,,\\
\hat{h} & = \so(N)_{N-1}\,,
\end{aligned}
\qquad\qquad
\begin{aligned}[c]
\hat{g}' & = \so(N-1)_{N} \oplus \so(N-1)_1\,,\\
\hat{h}' & = \so(N-1)_{N+1}\,.
\end{aligned}
\end{equation}
Using the embeddings 
\be
\so(N)_{N-2}\subset\so\big(\tfrac{N(N-1)}{2}\big)_1 \textrm{~~~and~~~} \so(N-1)_{N+1}\subset \so \big (\tfrac{(N-2)(N+1)}{2} \big)_1\,,
\ee
listed in Table~(\ref{t:1}), we find that
\begin{equation}
\hat{g}\oplus\hat{h}' \subset \so\big(\tfrac{N(N-1)}{2}\big)_1 \oplus \so \big (\tfrac{(N-2)(N+1)}{2} \big)_1 \oplus \so(N)_1\,.
\end{equation}
Further using the conformal embedding $\so(M)_1\oplus\so(N)_1\subset\so(M+N)_1$ one can establish that
\begin{equation}
\hat{g}\oplus\hat{h}'\subset\so(N^2-1)_1
\end{equation}
is also a conformal embedding.
By a similar argument one can see that 
\begin{equation}
\hat{g}'\oplus\hat{h}\subset\so(N^2-1)_1
\end{equation}
is also a conformal embedding. Therefore the two $SO$-coset series of \equ{SOSeries} are $T$-equivalent. 

Series~I and Series~II of $SU$-cosets given in \equ{SUSeries} can be shown to be $T$-equivalent with an additional factor of $\mathfrak{u}(1)_1$. To be more precise, it can be shown that the two cosets 
\begin{equation}
\begin{aligned}[c]
\frac{\su(N)_{N+1}\oplus\su(N)_1\oplus\mathfrak{u}(1)_1}{\su(N)_{N+2}} & 
\end{aligned}
\quad\text{and}\quad
\begin{aligned}[c]
\frac{\su(N+1)_{N-1}\oplus\su(N+1)_1\oplus\mathfrak{u}(1)_1}{\su(N+1)_{N}} & 
\end{aligned}
\end{equation}
are $T$-equivalent. For these two cosets we have:
\begin{equation}
\begin{aligned}[c]
\hat{g} & = \su(N)_{N+1} \oplus \su(N)_1\oplus\mathfrak{u}(1)_1\,,\\
\hat{h} & = \su(N)_{N+2}\,,
\end{aligned}
\quad\quad
\begin{aligned}[c]
\hat{g}' & = \su(N+1)_{N-1} \oplus \su(N+1)_1\oplus\mathfrak{u}(1)_1\,,\\
\hat{h}' & = \su(N+1)_{N}\,.
\end{aligned}
\label{modifSUcosets}
\end{equation}
Making use of the conformal embeddings
\be
 \su(N)_{N+1}\oplus\su(N+1)_N\subset\su(N(N+1))_1~~\textrm{and}~~ \su(N)_1\oplus\mathfrak{u}(1)_1\subset\su(N+1)_1
\ee
we see that 
\begin{equation}
\hat{g}\oplus\hat{h}'\subset\su(N(N+1))_1\oplus\su(N+1)_1
\end{equation}
is a conformal embedding. Similarly, using the conformal embeddings:
\begin{alignat}{1}
 \su(N)_{N+2}&\subset\su(N(N+1)/2)_1\,, \nn \\
 \su(N+1)_{N-1}&\subset\su(N(N+1)/2)_1\,, \nn \\
\su(M)_1\oplus\su(N)_1\oplus\mathfrak{u}(1)_1&\subset\su(M+N)_1
 \end{alignat}
we can see that 
\begin{equation}
\hat{g}'\oplus\hat{h}\subset\su(N(N+1))_1\oplus\su(N+1)_1
\end{equation}
is also a conformal embedding. Hence the Series~I and Series~II coset models of \equ{modifSUcosets} are $T$-equivalent, albeit in a weaker form than the Series~V and Series~VI cosets.

\subsection{Proposition \label{prop}}
The branching functions of T-equivalent coset models are, in general, related in some way \cite{Bowcock, Altschuler}. This can be seen in the context of Series~I and Series~II coset models as follows:  As shown in the last section, combinations of the constituent groups of the coset models are embedded in $SU(N(N+1))_1 \oplus SU(N+1)_1$. A character of $\mathfrak{su}(N(N+1))_1 \oplus \mathfrak{su}(N+1)_1$  can be expanded in terms of the characters of $\hat g$ and $\hat h'$ on one hand, and $\hat g'$ and $\hat h$ on the other, with $\hat g,\hat h', \hat g', \hat h$ defined in \equ{modifSUcosets}. These expansions results in a relation between the branching functions of $\hat g/\hat h$ and of $\hat g'/\hat h'$. To actually find the relation between the branching function of the Series~I and Series~II coset models using this procedure, one has to disentangle the characters of the additional $U(1)$ from the equations obtained. This is somewhat complex and we leave this to future work. We note that the Series~V and Series~VI coset models do not have this problem and hence it should be straightforward to derive the relations between their branching functions. Our aim in this section is to write down the identities, involving the branching functions of the Series~I and Series~II cosets, that we expect to be true based on the numerical evidence for low-lying $N$ values.

We first show that an exceptional-type partition function of the  $SU(N)_{N+2}$ WZW model is equal to a partition function of the $SU(N+1)_{N-1}$ model. 
The character $X_\l$ of the $SU(\frac{N(N+1)}{2})_1$ model, corresponding to a primary field of weight $\l$, can be split in two ways:
\beq
\label{Cexp1}
	X_\l &= c_{\l\d} \Phi^{\I}_{\d}, \\
\label{Cexp2}
	X_\l &= d_{\l\b} \Phi^{\I\I}_{\b}.
\eeq
Here $\Phi^{\I}_\d$ is the $SU(N)_{N+2}$ character while $\Phi^{\I\I}_{\b}$ is the $SU(N+1)_{N-1}$ character, and a double index implies summation. The coefficients $c_{\l\d} $ and  $d_{\l\b} $ are independent of $\tau$. The above results in the character identities
\be \label{wzwIden}
	c_{\l\d} \Phi^{\I}_{\d} = d_{\l\b} \Phi^{\I\I}_{\b}.
\ee
The number of these identities corresponds to the number of characters $X_\l$ of $SU(\frac{N(N+1)}{2})_1$, which is $N(N+1)/2$, equal to the number of primary fields.

The diagonal partition function of the $SU(\frac{N(N+1)}{2})_1$ model is $X_\l \overline{X}_\l$. Using the identities in \equ{Cexp1}, this results in the exceptional-type partition function of $SU(N)$ at level ${N+2}$
\be
	{\cal Z}^I_E = c_{\l\d} c_{\l\b} \Phi^{\I}_{\d} \overline{ \Phi}^{\I}_{\b}.
\ee
Similarly, there exists a partition function for $SU(N+1)$ at level $N-1$, given by 
\be
	{\cal Z}^{\I\I}_E = d_{\l\d} d_{\l\b} \Phi^{\I\I}_{\d} \overline{ \Phi}^{\I\I}_{\b}.
\ee
Because of the identities in \equ{wzwIden}, there exists the corresponding identity 
\be
{\cal Z}^I_E ={\cal Z}^{\I\I}_E. 
\ee

Based on the numerical evidence for $N=3,4,5$, our claim is that the branching functions $b^{\I}_{\b\b'}$ of Series~I and $b^{\I\I}_{\d\d'}$ of Series~II obey the following identities:
\be \label{branchEquiv}
	c_{\l \d} b^{\I}_{\d \a} = d_{\l' \b} b^{\I\I}_{\a'\b}.
\ee
The coefficients $c_{\l\d} $ and $d_{\l'\b}$ that appear above are the same as in \equ{wzwIden}. We propose that there is a one-to-one map between the indices $(\l,\a)$ appearing on the L.H.S and the indices $(\l',\a')$ appearing on the R.H.S. Below we show that this statement is at least compatible with a counting of the primary fields of the coset models that appear on either side. The indices $\l$ and $\l'$ both label the weights of $SU(\frac{N(N+1)}{2})_1$. The index $\a$ labels a weight of $SU(N)_{N+1}$ and the index $\a'$ labels a weight of $SU(N+1)_N$. Even though, the number of primary fields of these two WZW models are different, a one-to-one map between them can still exist in the coset context, because we also need to take field identification into account. For a $SU(M)_k$ WZW model, the number of primary fields is given by $\frac{(k+M-1)!}{k! (M-1)!}$. The number of primary fields of the WZW  $SU(N)_{N+1}$ model is, thus, $\frac{(2N)!}{(N-1)!(N+1)!}$.  We can choose to restrict the set of primary fields of the Series~II coset such that we include primary fields with no restriction on the weights of the $SU(N)_{N+2}$ factor, but only include $1/N$ of the weights of the $SU(N)_{N+1}$ factor. The index $\a$ then has count $\frac{(2N)!}{(N)!(N+1)!}$. Similarly, for Series~III, we can restrict the set of primaries such that we keep fields with any weight of the  $SU(N+1)_{N-1}$ factor but with $1/(N+1)$ weights of the $SU(N+1)_{N}$ factor. The index $\a'$ again has count $\frac{(2N)!}{(N)!(N+1)!}$. The total number of identities is $\frac{(2N)!}{2 (N-1)!(N)!}$. These concepts are illustrated for the $N=3$ case in Appendix~(\ref{seriesEx}).

Note the fact that the count of the index $\a$ is the same as the count of the index $\a'$ (at the coset level) is actually a reflection of the level-rank duality between $SU(N)_{N+1}$ and $SU(N+1)_{N}$. In fact, we can fix the set of elements $\a'$, given the set of elements $\a$. For a representation $\a$ of $SU(N)_{N+1}$, the corresponding representation $\a'$ of $SU(N+1)_{N}$ can be determined by exchanging rows for columns in the Young tableau for $\a$ \cite{Naculich:1990}. For the $N=3$ case, for instance, this maps the set in \equ{alphawts1} to the set in \equ{alphawts2}.

The reason the index $\l$ in \equ{branchEquiv} is not simply equal to $\lambda'$ is because the identities at the coset level are dependent on the conformal dimension of the coset primary fields as stated at the end of \sct{N3}. Formally, let us denote the weights associated with the coset primary field with non-zero coefficient (that is, non-zero $c_{\l \d}$ and $d_{\l' \b}$) and minimum conformal weight by $(\l; \a, \delta_\textrm{min})$ on the L.H.S of \equ{branchEquiv} and by $(\l'; \a',\beta_\textrm{min})$ on the R.H.S.
Then, a identity of the form in \equ{branchEquiv} can only work, if the conformal dimension of the fields labelled by these weights are equal, that is:
 \be
   h[(\l; \a, \delta_\textrm{min})] = h[(\l'; \a',\beta_\textrm{min})].  
 \ee
The conformal dimension $h$ depends on the weights of the constituent WZW models as in \equ{cdimension}. It is not always necessary that the weight $\b_\textrm{min}$ that solves the above equation for fixed $\a$ and $\delta_\textrm{min}$ is present in the branching of the character of the specific representation $\l$ into representations of $SU(N+1)_{N-1}$. This is again illustrated in Appendix~\ref{seriesEx}. 

The statement of equivalence of partition functions for the Series~I and Series~II coset models is as follows. If the identities in \equ{branchEquiv} are satisfied, it will imply that the $AE$ partition function of Series~I given by
\be
	{\cal Z}^{\I}_{AE} = c_{\l\d} c_{\l\b} b^{\I}_{\d \a} \overline{ b}^{\I}_{\b \a} 
\ee
and the $EA$ partition function of Series~II given by
\be
	{\cal Z}^{\I\I}_{EA} =  d_{\l\d'} d_{\l\b'} b^{\I\I}_{\a' \d'} \overline{ b}^{\I\I}_{ \a' \b'} 
\ee
are equal. To complete the proof one needs to prove the identities in \equ{branchEquiv}.

\section{A $\mathcal{N}=1$ even-spin CFT algebra and its bulk dual \label{BulkDuals}}

Out of the conformal embeddings listed in Table~(\ref{t:1}), the most interesting cases are the embeddings in the first row which result in supersymmetric coset models. Of these, the coset model corresponding to the case $SU(N)_N \subset SO(N^2-1)_1$ has already appeared in the literature \cite{Beccaria}. In this section, we study the coset model
\be \label{dncoset}
	\frac{SO(2 N)_{2N-2} \otimes SO(2N)_1}{SO(2N)_{2N-1}}\,,
\ee 
which we refer to as the $D_N$ coset model at fixed level. This coset model is a special case of a $D_N$ coset model at general level $k$:
\be \label{gdncoset}
	\frac{SO(2N)_{k} \otimes SO(2N)_1}{SO(2N)_{k+1}}
\ee
As stated earlier, the uniqueness of the coset model in \equ{dncoset} stems from the fact that when the level $k=2N-2$, exceptional invariants with extended symmetry algebras appear.

We quickly review the coset model in \equ{gdncoset} and its corresponding bulk dual which is a higher-spin theory with gauge group $\textrm{hs}^e(\lambda)$.  For details, the reader is referred to \cite{Gaberdiel:2011nt}. The spectrum of the coset diagonal invariant consists of representations specified by $(\L_+,\L_-)$, where $\L_+$ and $\L_-$ are highest weight representations of  $SO(2N)_k$ and $SO(2N)_{k+1}$ respectively. The vacuum sector of the theory corresponds to both $\L_+$ and $\L_-$ being  identity representations. For the diagonal invariant, the vacuum sector determines the spin content of the symmetry algebra of the coset model. This algebra has been shown to be a bosonic algebra containing currents of spin $2,4,6,\cdots, \infty$, known as $\W^e_\infty(\lambda)$, which is also equivalent to the asymptotic symmetry algebra of the bulk theory. CFT representations of the form $(\L_+,0)$ map to a real scalar in the bulk, while representations of the form $(0,\L_-)$ map to conical defects \cite{Castro:2011}. In the 't Hooft limit, certain states decouple on the CFT side and the resulting partition function is exactly equal to the bulk thermal partition function. 

Our goal in this section is to identify a bulk dual for the $EA$ invariant of the coset model in \equ{dncoset}. For this purpose, we first compute the exact symmetry algebra of this invariant in \sct{dnfixed}. We propose the bulk dual in \sct{dndual} and show that its partition function agrees with the CFT theory. This serves as a proof-of-concept for the existence of the bulk dual of this coset model. We leave the finer details of this duality, for example, matching the non-vacuum sector of the CFT partition function with a matter sector in the bulk, to future work.
\subsection{The $D_N$ coset at a fixed level \label{dnfixed}}
As stated previously, the extended algebra for the coset model in \equ{dncoset} is the algebra associated with the diagonal invariant of the coset
\be \label{sdncoset}
	\frac{SO(M)_1 \otimes SO(2N)_1}{SO(2N)_{2N-1}}\\,
\ee
where $M=N(2N-1)$.
From the calculation in Sec.~(\ref{exs}), we expect this algebra to be supersymmetric.  One way to find the symmetry algebra of this coset model is to calculate its vacuum branching function, which we denote by $\B_{(0;0)}$, directly. However, it is much easier to compute the branching function of the coset model in \equ{dncoset} and then sum over the relevant sectors to get $\B_{(0;0)}$. To determine which sectors go into the summation, we find the branching rule for the vacuum sector of the coset model in \equ{sdncoset} in \sct{br}. Then in \sct{vpt} we find the symmetry algebra. 
\subsubsection{Branching rules \label{br}}
The diagonal partition function for the coset in \equ{sdncoset} can be rearranged in manifestly supersymmetric form. In the following, we denote a representation of any coset model as $(\L_+,\L_-)$ --- we hope it is clear from context which coset model it is a representation of. The vacuum sector ${(0;0)}$ for the supersymmetric diagonal invariant is given by a sum of the $(0;0)$ and $(\textrm{v};0)$ sectors of the coset model in \equ{sdncoset}. These sectors, further, decompose into sectors of the coset model in \equ{dncoset}, according to the appropriate branching rule.

The branching rule for the $(0;0)$ and $(\textrm{v};0)$ sectors of the coset model in \equ{sdncoset} is determined solely by the branching rule of the $0$ and $\textrm{v}$ representations of the WZW group $SO(M)_1$ into representations of $SO(2N)_{2N-2}$.  Since we are looking at a $EA$ type invariant, the representation associated with the $SO(2N)_{2N-1}$ group remains fixed on both sides of the branching rule for the coset model, being in this case the identity representation.

The branching rule for decomposing the $SO(M)_1$ representations into representations of $SO(2N)_{2N-2}$ can be worked out explicitly for small values of $N$, using the method in Ref.~\cite{Kac:1988}. For example, for $N=3$, the branching rule for the vector and vacuum representations of $SO(15)_1$ into representations of $SO(6)_4$ is:
\begin{align}
	(0^7) &\longrightarrow (0, 0, 0) \oplus  (1, 0, 2) \oplus (1, 2, 0) \oplus (4, 0, 0)\,, \\
	(1,0^6) &\longrightarrow (0, 0, 4) \oplus  (0, 1, 1) \oplus (0, 4, 0) \oplus (2, 1, 1)	\,.
\end{align}
As a general rule, it is a necessary (but not a sufficient) condition that only those representations $\L$ can appear in the branching of a representation $\Pi$, whose conformal weight differs from the conformal weight of $\Pi$ by integers. The weights of the vacuum and vector representations of $SO(M)_1$ are $0$ and $\frac{1}{2}$ respectively. Therefore, to find the set of weights of $SO(2N)_{2N-2}$, that can appear in the branching of these representations we need to find the set of weights whose conformal dimension $h_L$ is an integer or a half-integer.
In the large $N$ limit, the conformal dimension of a representation $\L$ of $SO(2N)_{2N-2}$ is given by
\be \label{confdim}
	h_\L = \frac{ C_N(\L)}{4N-4} \cong \frac{B(\L)}{4} + \frac{D(\L)+B(\L)}{2(4N-4)} \,.
\ee
Here, $B(\L)$ is the number of boxes in the Young tableaux of the representation $\L$ and $D(\L)$ is defined in \equ{A.17}.  For $h_L$ to be integer or half-integer, both the first and second terms in \equ{confdim} should be separately integer and half-integer. Therefore, the condition that a weight appear in the branching of $\B_{(0;0)}$, in the infinite $N$ limit, is that $B$ is an even non-negative integer and that  $D(\L)+B(\L)$ be a multiple of $4N-4$ (including zero). However, since in the large $N$ limit we only include those weights in the partition function whose number of boxes $B(\L)$ is finite,  and $D(\L)$ being of $O(1)$, the second condition reduces to $D(\L) +B(\L)=0$. 
We now examine what Young tableau satisfy these requirements. This is best seen in the Frobenius notation for these diagrams, which is reviewed in Appendix~(\ref{Fso}). In the Frobenius notation, $B(\L)$ and $D(\L)$ are given by Eqs.~(\ref{BF}) and (\ref{DF}) respectively. Therefore, the condition $D(\L)+B(\L)=0$ becomes
\be \label{cond2}
	 \sum_{i = 1}^{d} \(a_i + b_i + 1\) =  \sum_{i=1}^d \(b_i + \tfrac{1}{2}\)^2 - \sum_{i = 1}^d \(a_i + \tfrac{1}{2}\)^2 \,.
\ee
It is easy to see that a representation with Frobenius coordinates of the generic form
\begin{equation} 
\left( \begin{array}{cccc}
a_1 & a_2 &\dots & a_d \\
a_1+1 & a_2+1 &\dots & a_d+1 
\end{array} \right)\,,
\label{Frob}
\end{equation} 
will always satisfy the condition in \equ{cond2} as well as the condition that $B \in 2 \mathbb Z_{\geq 0}$.  We denote this set of Young diagrams by $\Sigma$. Note that the identity representation also belongs to this set $\Sigma$.
In addition, there are some representations that are not of the simple form in \equ{Frob}, but are solutions of the \equ{cond2}. However, one can verify that they do not appear in the branching of $\B_{(0,0)}$ at finite $N$ and, therefore, also do not appear in the infinite $N$ limit. It can be explicitly checked that the branching rule for the vacuum sector of $SO(M)_1$, denoted as Vac below, for small values of $N$ takes the following form in Frobenius notation:
\be
	\textrm{Vac}  \rightarrow  \textrm{Vac} + \left( \begin{array}{cccc}
1 \\ 2
\end{array} \right)\, + \, \left( \begin{array}{cccc}
2 & 1  \\
3 & 2
\end{array} \right)\, + \cdots
\ee
if we retain only the representations that appear in the infinite $N$ limit. The branching of the vector representation takes a similar form.

\subsubsection{The vacuum partition function \label{vpt}}
The vacuum branching function of the coset model in \equ{sdncoset} is given by
\be \label{susybr}
	\B_{(0;0)} =  \sum_{\L_+ \in \Sigma} b_{(\L_+,0)}\,.
\ee
The branching function $b_{(\L_+,0)}$ of the coset in \equ{dncoset} is worked out in Appendix~(\ref{gco}), in the 't Hooft limit.  In the following, we denote the Young tableaux associated with a representation $\Lambda$ as $Y(\L)$, with the transpose denoted by $Y^T(\L)$.  Then, the branching function can be written as 
\be\label{generalBranching}
b_{( \Lambda_{+}; 0 )} (q) \cong q^{-\frac{c}{24}}\, M^e (q) 
P^+_{Y^T(\L_+)} (q) \,.
\ee
Here, $M^e (q)$ is the modified even MacMahon function:
\be\label{MMe}
M^e (q) \equiv \prod_{\substack{s = 2 \\ s \, \mathrm{even}}}^\infty \; \prod_{n = s}^\infty 
\frac{1}{1 - q^n } \,.
\ee
The  $P_{Y(\L)}^{\pm}(q)$ are the modified Schur functions
\begin{eqnarray}\label{Schur}
P_{Y(\L)}^\pm(q) & = & q^{\pm\frac{\lambda}{2} B(Y)} \, \textrm{ch}_\L(U_h) \,,
\end{eqnarray}
where $\textrm{ch}_\L(U_h)$ is the Schur polynomial defined as
\be	
	\textrm{ch}_\L(U_h) = \sum_{T\in \textrm{Tab}_\L} \prod_{i \in T} q^{h+i}\,,
\ee
with $U_h$ a diagonal matrix having matrix elements ${(U_h)}_{ii}=q^{i + h}$. The sum is over a filling of the boxes of a semistandard Young tableau of shape $\L$ with integers $i \geq 0$.
When the level $k=2N-2$, the coupling $\lambda=\frac{1}{2}$. Using the identity:
\be
	q^{\frac{1}{4}B(Y)} \textrm{ch}_\L(U_{\frac{1}{2}}) = \textrm{ch}_\L(U_{\frac{3}{4}})\,,
\ee
the branching function becomes
\be
	b_{( \Lambda_{+}; 0 )} \cong q^{-\frac{c}{24}}\, M^e (q)  \,  \textrm{ch}_{\L^{T}}(U_{\frac{3}{4}})\,.
\ee
 To get the vacuum character of the supersymmetric theory we sum over all representations that belong to the set $\Sigma$, as stated in \equ{susybr}. The vacuum character is then
 \be\label{susyVacuum}
 	\B_{(0;0)}(q)= \sum_{\L \in \Sigma} q^{-\frac{c}{24}}\, M^e (q)  \,  \textrm{ch}_{\L^{T}}(U_{\frac{3}{4}})\,.
 \ee 
To extract the higher spin algebra from the vacuum character, we make use of a Littlewood identity which appears, for example, in \cite{littlewood1,littlewood2}. (It is the identity in Eq.~(4) in Ref.~\cite{littlewood1}.) The identity is
\be
	\sum_{\L \in \Sigma^T} \textrm{ch}_{\L}(U_{h}) = \prod_{i\geq j} (1+ q^{i+j+2 h})\,,
\ee
where the variable $j$ runs from $0$ to $\infty$.  Here, $\Sigma^T$ is the set of representations that are transpose of the representations in $\Sigma$ and in Frobenius notation are of the form:
\be
\left( \begin{array}{cccc}
a_1+1 & a_2+1 &\dots & a_d+1\\
a_1 & a_2 &\dots & a_d 
\end{array} \right)
\ee
and also include the identity representation. The vacuum character then becomes:
\be
	\B_{(0;0)}(q)= q^{-\frac{c}{24}}\, M^e (q)  \prod_{i\geq j} (1+ q^{i+j+3/2}) = q^{-\frac{c}{24}}\, M^e (q)   \prod_{\substack{s = 1 \\ s \, \mathrm{odd}}}^\infty \; \prod_{n = s}^\infty 
(1 + q^{n+\frac{1}{2}})\,.
\ee
The spin content of the vacuum algebra is, therefore, 
\be \label{soalgebra}
	(\tfrac{3}{2},2),\, (\tfrac{7}{2},4),\, (\tfrac{11}{2}, 6),\, \cdots,\, (2k + \tfrac{3}{2},2k + 2),\, \cdots \,,
\ee
where $k =0,1,2, \cdots$. This is a $\mathcal{N}=1$ supersymmetric algebra, which we denote by $s\W^e_\infty$.

\subsection{Bulk dual \label{dndual}}
In this section, we show that there exists a consistent higher-spin theory in the bulk with algebra corresponding to the spectrum in  \equ{soalgebra}.  An algebra of this form first appeared in Ref.~\cite{vas1}, for bulk dimension $D=4$. 
In that paper, the authors showed that the $\mathcal{N}=1$ supersymmetric $\textrm{shs}_\rho(1)$ algebra, where $\rho$ is a parameter with values either $0$ or $1$, contains subalgebras with fields having spin 
\be
	s= 2k+2 \textrm{~~~and~~~} s=2k + \tfrac{3}{2} + \alpha\,
\ee
where $\alpha$ is either $0$ or $1$. These algebras are denoted as $\textrm{shs}(1|\a)\equiv \textrm{shs}_\rho(1|\a)$, since they are independent of $\rho$. For both values of $\a$, these algebras are superalgebras, but the $\alpha=1$ case is not supersymmetric. Subsequently, in Refs.~\cite{vas3} and \cite{blencowe}, it was shown that the same structure exists in $D=3$. Note that the field content of the $\textrm{shs}(1|0)$ algebra coincides with that of \equ{soalgebra}.

The general higher-spin theory in the bulk is a $\N=2$ supersymmetric theory which has a free parameter $\m$ related to the masses of the matter fields. This is known as the Prokushkin-Vasiliev theory and has a gauge group corresponding to an algebra known as $\textrm{shs}(\m)$,  with the asymptotic symmetry algebra of the theory being $s\W_\infty$. 
In \sct{shstrunc}, we list possible truncations of the $\textrm{shs}(\mu)$ algebra, from the viewpoint of the higher-spin algebra being a wedge algebra of the $\N=2$ $s\W_\infty$ algebra and show that the $\textrm{shs}(1|0)$ algebra is an allowed truncation. In \sct{shstrunc} we show that there exists a truncation of the full $\N=2$ Prokushkin-Vasiliev theory associated with this algebra, following Ref.~\cite{Prokushkin:1998}.

\subsubsection{Truncation of $\textrm{shs}(\mu)$ algebra \label{shstrunc}}
In this section we sketch how the $\textrm{shs}(1|0)$ algebra can be constructed as a subalgebra of the $\N=2$ $\textrm{shs}(\mu)$ algebra when $\mu=1/2$.  In fact, at this special point, the $\N=2$ $\textrm{shs}(\mu)$ algebra has a number of subalgebras \cite{vas2}. 

For general $\mu$, the standard method to construct the $\textrm{shs}(\mu)$ algebra is to factor the universal enveloping algebra of $\mathfrak{osp}(1,2)$ by an ideal. In detail,
\be
	\textrm{shs}(\mu) \oplus \mathbb{C} = \frac{U(\mathfrak{osp}(1,2))}{\langle C_{\mathfrak{osp}(1,2) }- \frac{1}{4} \mu (\mu-1) \bf 1 \rangle}\,,
\ee
where $C_{\mathfrak{osp}(1,2)}$ is the quadratic casimir of $\mathfrak{osp}(1,2)$.
The generators of the $\textrm{shs}(\mu)$ algebra can be constructed \cite{vas2, Beccaria} in terms of $V^{(s)\pm}_m$ defined as
\be \label{shsG}
	V^{(s)\pm}_m = \ty_{(\a_1\cdots} \ty_{\a_n)} (1\pm Q)\,.
\ee
where the operators $y_\a$ obey the algebra
\be \label{ocom}
	\left [\ty_\a, \ty_\b \right]=2i\e_{\a\b}\{1+(2\mu - 1) Q\}\,,~~~~~~~\{Q, \ty_\a\}=0\,,~~~~~~~Q^2=\mathbf{1}\,.
\ee
Changing the basis to
\be
	W^{(s)+}_m \equiv V^{(s)+}_m + V^{(s)-}_m,
\ee
the bosonic generators of $\mathfrak{osp}(1,2)$, which is a subalgebra of $\textrm{shs}(\mu)$, can be written as follows:
\be
	L_0= \tfrac{i}{8}(W^{(2)+}_0)\,,\,\,\,\,L_{+1}=\tfrac{i}{4}(W^{(2)+}_{+1}) \,,\,\,\,\,L_{-1}=\tfrac{i}{4}(W^{(2)+}_{-1})\,.
\ee
The fermionic generators are
\be
	G_{+\frac{1}{2}} =\tfrac{1}{4}\,e^{-\frac{i \pi}{4}} (W^{(\frac{3}{2})+}_{+\frac{1}{2}}) \,,\,\,\,\, G_{-\frac{1}{2}} =\tfrac{1}{4}\,e^{-\frac{i \pi}{4}} (W^{(\frac{3}{2})+}_{-\frac{1}{2}}) \,.
\ee
As is apparent, there is a second set of bosonic and fermionic generators that can be constructed as $W^{(s)-}_m \equiv V^{(s)+}_m - V^{(s)-}_m$. In fact, the $\textrm{shs}(\mu)$ algebra can also be constructed as a quotient of the universal enveloping algebra of $\mathfrak{osp}(2|2)$. The algebra is, therefore, a $\N=2$ supersymmetric algebra with field content:
\be
	(1,\tfrac{3}{2},\tfrac{3}{2},2)\,,\,\, (2,\tfrac{5}{2},\tfrac{5}{2},3)\,,\,\,(3,\tfrac{7}{2},\tfrac{7}{2},4)\,\cdots
\ee
The $\textrm{shs}(\mu)$ algebra is the wedge algebra of the $\N=2$ $s\W_\infty$ algebra in the $c\rightarrow \infty$ limit. The generators of the $s\W_\infty$ algebra are $L_{n}^{(\tilde s)\pm}$ 
and $G_{n}^{(\ts)\pm}$, where $\ts$ is a integer obeying $\ts\geq 2$, and $L_{n}^{(1)-}$. The spin of the operator $L_{n}^{(\ts)\pm}$  is $s=\ts$ while the spin of $G_{n}^{(\ts)\pm}$ is $s=\ts-\frac{1}{2}$. Therefore, the algebra consists of the supermultiplets $(\ts,\ts-\frac{1}{2})$ corresponding to the generators $(L_{n}^{(\ts)\pm},G_{n}^{(\ts)\pm})$.
For this algebra one can implement an automorphism \cite{vas2}, such that the generators transform as follows:
\be \label{autom}
	L_\mu^{(\ts)\pm} \rightarrow \pm (-1)^{\ts-1} L_{1-\mu}^{(\ts)\pm} \,,\,\,\,\,\, G_\mu^{(\ts)\pm} \rightarrow i (-1)^{\ts-1} G_{1-\mu}^{(\ts)\pm}\,.
\ee
Note that when $\mu$ takes the value $1/2$, any generator maps to itself. The structure constants of this algebra take the form:
\be \label{scsymmetry}
	f^u_{st}(\m)=F^u_{st}(\m) + (-1)^{[-u] + 4(s+u)(t+u)} F^u_{st}(1- \m)\,,
\ee
where $F^u_{st}(\m)$ is a function of $\m$ and the spins $s,t,u$ and 
\be
    [u] \equiv 
\begin{cases}
u \textrm{~if~}u \in \mathbb Z\,, \\
 u-1/2 \textrm{~if~}u  \in \mathbb Z + 1/2
 \end{cases}
\ee
For $\m=1/2$, many of the structure constants vanish and the algebra closes for a reduced set of generators. It can be shown, using, for example, \equ{scsymmetry}, that the algebra can be consistently truncated to retain the generators $L_\mu^{(\ts)+}$ with $\ts$ even, $L_\mu^{(\ts)-}$ with $\ts$ odd and the generators $G_\mu^{(\ts)\pm}$ with $\ts$ either even or odd.  If one retains the $G_\mu^{(\ts)\pm}$ with odd $\ts$, one gets an algebra with no supersymmetry with the following field content:
\be
	1,2,\tfrac{5}{2},\tfrac{5}{2},3,4,\tfrac{9}{2},\tfrac{9}{2}\,\cdots\,.
\ee
On the other hand, retaining the $G_\mu^{(\ts)\pm}$ with even $\ts$ preserves the $\N=2$ symmetry of the original theory, with the field content
\be
\label{n2trun}
	(1,\tfrac{3}{2},\tfrac{3}{2},2)\,,\,\,\,\,(3,\tfrac{7}{2},\tfrac{7}{2},4)\,\cdots\,.
\ee
Restricting the above $\W$-algebra to its wedge modes, one should get a $\N=2$ supersymmetric higher-spin algebra which we denote by $\textrm{shs}(2|0)$.  As for the $\W$-algebra, for the $\textrm{shs}$ algebra, we can retain the generators $W^{(s)+}_m $  with $s = 2k + 2$ and $s=2k+\frac{3}{2}$ and the generators $W^{(s)-}_m$  with $s = 2k+1$ and $s=2k +\frac{3}{2}$, where $k\in {0,1,2,\cdots}$. 

Other truncations of the $\textrm{shs}(\mu)$ algebra are possible. In \equ{ocom}, for $\mu=1/2$, the commutators of the $\ty_\a$ become independent of $Q$. As a consequence, one can choose only to retain the $W^{(s)+}$ generators \cite{Beccaria}. In this case, the supersymmetry, reduces from $\N=2$ to $\N=1$ and this algebra is known as the $\textrm{shs}(1|2)$ algebra. The automorphism in \equ{autom} still applies, and as for the $\N=2$ case, only the generators $W^{(s)+}_m $  with $s = 2k+2$ and $s=2k+\frac{3}{2}$ can be retained to give the truncated algebra $\textrm{shs}(1|0)$.

\subsubsection{Truncation of Prokushkin-Vasiliev theory}
We now show that this $\textrm{shs}(1|0)$ algebra is associated with a truncation of the $\N=2$ higher-spin Prokushkin-Vasiliev theory. This construction appears in Ref.~\cite{Prokushkin:1998} and we review the salient features here.

The field equations of the higher-spin theory are written in terms of the functions $W_\mu$, a space-time $1$-form, and  $B $ and  $S_\a$  which are space-time $0$-forms, with $\a$ a spinor index taking values $1,2$. 
These are generating functions, with $W_\mu$ being the generator of the higher-spin gauge fields, $B$ the generator of matter fields, while $S_\a$ is for auxiliary fields. 
These generators are functions of the space-time coordinates $x_\mu$ and the auxiliary variables $(z_\alpha , y_\alpha ; \psi_{1,2}, Q , \rho)$.  Here, $z_\alpha , y_\alpha$ are commuting bosonic twistor variables, while the $(\psi_1,\psi_2)$ and $(Q,\rho)$ are two sets of Clifford elements.
The generating functions are expanded as
\begin{align}
\label{genfun}
 A (z , y ; \psi_{1,2} , Q , \rho | x) = \sum_{B,C,D,E = 0}^1 \sum_{m,n = 0}^\infty
  A^{BCDE}_{\alpha_1 , \ldots , \alpha_m , \beta_1 , \ldots , \beta_n}
  Q^B \rho ^C \psi_1^D \psi ^E_2 z^{\alpha_1} \ldots  z^{\alpha_m}
y^{\beta_1} \ldots y^{\beta_n} ~.
\end{align}
The coefficients $ A^{BCDE}_{\alpha_1 , \ldots , \alpha_m , \beta_1 , \ldots , \beta_n}$ carry spin $s=(n+m)/2 +1$ and commute with the generating elements $z_\alpha , y_\alpha , \psi_{1,2}, Q , \rho$.
These generating functions obey a system of equations and we will work with a vacuum solution of these equations.
We choose a zero-order vacuum solution for the matter field: $B = \nu$, where $\n$ is a constant. Simultaneously, $S_\alpha$ can be chosen to equal $S_{\alpha,0}^\text{sym}$, which is defined in Ref.~\cite{Prokushkin:1998}.
For this choice of $B$,  the vacuum solution $W =W_0$ can be shown to depend only on $(\tilde y_\alpha ; \psi_1,Q)$, where $\tilde y_\alpha$ is called the ``deformed-oscillator'' and is equal to $\tilde y_\alpha^\text{sym}$ of Ref.~\cite{Prokushkin:1998}.  The variables  $\tilde y_\alpha$  can be shown to obey the same form of commutation relations as the undeformed oscillators $y_\a$:
\begin{align}\label{bcom}
 [ \tilde y_\alpha , \tilde y_\beta ] = 2 i \epsilon_{\alpha \beta} (1 + \nu Q) ~,
 \qquad \{ \tilde y_\alpha , Q \} = 0 ~.
\end{align}
By incorporating the variable $\psi_1$ into projection operators, physical fields $\A,\bar \A$ can be defined as functions of $(\tilde  y_\alpha; Q)$ only.
Defining $\A,\bar \A$ as
\begin{align}
 W_0 = - \frac{1 + \psi_1}{2} \A  - \frac{1 - \psi_1}{2} \bar \A ~,
\end{align}
we have the expansions:
\be
\label{Aexp}
 \A (\tilde y ;  Q ) = \sum_{B= 0}^1 \sum_{m = 0}^\infty
  \A^{B}_{\alpha_1 ,\alpha_2,  \ldots , \alpha_m} Q^B  \tilde y^{\alpha_1} \star \ty^{\alpha_2}\ldots \star   \tilde y^{\alpha_m}\,,
\ee
where $\star$ denotes the Moyal star product. Note that the commutators in \equ{bcom} are the same as in \equ{ocom}, with $\nu$ identified as $2\mu-1$. In fact, the bulk theory is a Chern-Simons theory for the algebra generated by $Q$ and $\ty_\alpha$, which is the $\textrm{shs}(\mu)$ algebra of the previous section.

We now look at the symmetries of the Prokushkin-Vasiliev theory. We define a map $\sigma$ by
\begin{align}
 \sigma [A(z,y;\psi_{1,2},Q,\rho)] = A^\text{rev} (-iz,iy;\psi_{1,2},Q,\rho)
\end{align}
where the order of all generating elements is reversed in $A^\text{rev}$. At the level of the expansion in \equ{Aexp}, this translates to reversing the order of $Q$ and $\tilde y_\a$. Also, the action of the map $\sigma$ on the vacuum solution $S_{\alpha,0}^\text{sym}$  is 
\be
 \sigma[ S_{\alpha,0}^\text{sym}] = - i S_{\alpha,0}^\text{sym} \,,
\ee
while
\be
	 \sigma[ \tilde y_\alpha] = i \tilde y_\alpha \,.
\ee
There is a Grassmann parity $\pi$ associated with coefficients $A^{BCDE}_{\a_1 \cdots \a_m \b_1 \cdots \b_n}$ in the expansion in \equ{genfun}. This is determined by the number of spinor indices as follows:
\begin{align}
\pi (W_{ \alpha_1 , \ldots , \alpha_m , \beta_1 , \ldots , \beta_n}  ) &=
 \tfrac12 ( 1 - (-1)^{|m+n|} ) ~, \nonumber \\
 \pi (B_{\alpha_1 , \ldots , \alpha_m , \beta_1 , \ldots , \beta_n}  )& =  \tfrac12 ( 1 - (-1)^{|m+n|} )  ~,  \nonumber \\
 \pi (S_{ \alpha_1 , \ldots , \alpha_m , \beta_1 , \ldots , \beta_n}  ) &=  \tfrac12 ( 1 - (-1)^{|m+n+1|} ) ~.
\end{align}
It can be shown that the transformation  
\begin{align}
\label{Esym}
\eta (W_\mu) = - i^{\pi (W)} \sigma (W_\mu) ~, \qquad
\eta (B) = i^\pi (B) \sigma (B) ~, \qquad
\eta (S_\alpha) = i^{\pi (S)+1} \sigma (S_\alpha) 
\end{align}
is a symmetry of the field equations.
Applying this transformation to the expansion in \equ{genfun}, and keeping only the terms that are invariant, results in the following generators: All terms with even spin survive corresponding to generators of the form $\tilde y_{\a_1} ... \tilde y_{\a_n}$ and 
$ Q\,\ty_{\a_1} ... \ty_{\a_n}$ where $n \in 4\mathbb Z_+ -  2$. A single set of fermionic generators survive corresponding to terms of the form $\tilde y_{\a_1} ... \tilde y_{\a_n}$ when $n \in 4\mathbb Z_+ - 3$ and $Q \,\tilde y_{\a_1} ... \tilde y_{\a_n}$ when $n \in 4\mathbb Z_+ - 1$.
This results in a  $\mathcal{N}=1$ supersymmetric theory, which is related to a CFT dual in Ref.~\cite{Creutzig:2012}. 

For the case $\nu=0$, additional symmetries appear. In this case, there is an involutive symmetry
\be
\label{Ksym}
	\zeta[W(Q)] = W(-Q), \,\,\, \zeta[S_\a(Q)]=S_\a(-Q), \,\,\, \zeta[B(Q)] = -B(-Q),
\ee
of the field equations of the higher-spin theory. Under this map, it is clear that only the set of generators of the $\N=2$ theory that are independent of $Q$ will survive, reducing the symmetry again from $\N=2$ to $\N=1$. This is the theory with algebra $\textrm{shs}(1|2)$. However, one can further truncate the system using the transformation in \equ{Esym}. The reduced set of generators will be of the form $\tilde y_{\a_1} ... \tilde y_{\a_n}$ where $n \in 4\mathbb Z_+ -2$ or $n \in 4\mathbb Z_+ - 3$, since the symmetry in \equ{Ksym} has already removed the $Q$-dependent generators. This is the algebra $\textrm{shs}(1|0)$ of the previous section. A non-abelian version of this theory first appeared in Ref.~\cite{vas3}.

For completeness we add that applying the transformations in Eqs.~(\ref{Esym}) and (\ref{Ksym}) simultaneously, and keeping only the generators invariant under these transformations, results in a bulk theory with an algebra having fields with spins listed in \equ{n2trun}.

\section{Discussion \label{discussion}}

The main result of this paper is the proposal of a new duality between an $EA$-type exceptional invariant of the orthogonal coset of \equ{dncoset} and a $\N=1$ higher-spin bulk theory that arises as a truncation of a $\N=2$ supersymmetric Vasiliev theory for the value of $\mu=1/2$.  This duality can also be thought of as being between the diagonal invariant of the coset in \equ{sdncoset} and the bulk theory. As evidence for this proposal, we found the vacuum partition function of this coset CFT and demonstrated that it agrees with the bulk spectrum, in the 't Hooft limit. 

To put this proposal on a firmer footing, there is more work that can be done. In particular, we haven't shown that the non-vacuum sector of the CFT partition function maps to the matter sector of the bulk theory. Further, we have not checked that the $\N=1$ $s\W_\infty^e$ algebra can be independently constructed by imposing the Jacobi identities. It would also be nice to verify that this algebra truncates to a finite algebra when the central charge is equal to the coset central charge at a given value of $N$.  One can also explicitly check whether the bosonic $\W_\infty^e$ algebra is a subalgebra of $s\W_\infty^e$ and whether, in turn, $s\W_\infty^e$ is a subalgebra of other algebras. 

The CFT modular invariant of \equ{dncoset} belongs to a class of invariants that have enhanced supersymmetry linked with the fact that they arise from conformal embeddings. The CFT invariant discussed in this paper and the non-diagonal invariant of Ref.~\cite{Beccaria} both belong to this class. Besides these, as we showed in \sct{EnumerateMIs}, one can also construct similar non-diagonal invariants for cosets with constituent groups of the $B_N$ and $C_N$ series. In this paper we have studied the $D_N$ coset exclusively, but cosets of the $B_N$ and $C_N$ series can also have bulk duals. We expect them to behave as $\N=1$ supersymmetric counterparts of the cosets studied in Ref.~\cite{Candu:2012}.

Although, in this paper we only discussed cases which have $\N=1$ supersymmetry, there is no reason why the same procedure cannot work to boost the supersymmetry of coset CFTs from $\N=1$ to $\N=2$ and from $\N=2$ to $\N=3$. Indeed, on the bulk side at $\mu=1/2$, as discussed in \sct{dndual}, the $\N=2$ theory has a number of truncations which either retain the $\N=2$ supersymmetry or reduce it to $\N=1$. At the same time, it has a number of extensions with enhanced supersymmetry \cite{Prokushkin:1998}. Clearly the $\m=1/2$ value is special in this regard. An open problem is the bulk dual for the Vasiliev theory with algebra $\N=2$ $\textrm{shs}(2|0)$. This theory only exists for $\nu=0$ or $\mu=1/2$, so it follows that the dual coset theory should be at a fixed level. Another question of interest concerns the coset:
\be
	\frac{SO(2N+1)_k \otimes SO(2N)_1}{SO(2N)_{k+1}}\,,
\ee
which was studied in Ref.~\cite{Creutzig:2012} and whose diagonal invariant is dual to a bulk theory with $\N=1$ supersymmetry, existing for all values of $\nu$. It would be interesting to check whether this coset has an invariant with enhanced supersymmetry for the value of level $k$ at which the $SO(2N+1)_k$ group is conformally embedded in a group of the form $SO(M)_1$. Finally, on the bulk side there also exist non-abelian counterparts of the truncated theories. It would be aesthetically satisfying to have CFT duals for these theories, on the lines of Refs.~\cite{Creutzig:2014,Creutzig:2013}. 

From the perspective of the CFT at fixed $\l=1/2$, as we did in \sct{EnumerateMIs}, one can construct all possible series of coset invariants that result from conformal embeddings and look for bulk duals for these. As we demonstrated in \sct{SeriesEquivalence}, the set of distinct CFT partition functions is smaller than the set of all possible partition functions because partition functions of different coset models turn out to be related. It would be good to have a complete proof of this equivalence. In this paper, we have focused on CFT modular invariants that have enhanced symmetry as compared to the diagonal modular invariant. However,  coset CFT theories typically also have non-diagonal modular invariants which do not have such an enhanced symmetry. Often, these invariants exist for all ranks and, for example, even values of the level $k$ which means a well-defined 't Hooft limit exists. It would be interesting to look at bulk duals for these kind of non-diagonal invariants, as well.


\begin{acknowledgements}
We thank Rajesh Gopakumar for inspiration and useful suggestions. We thank Matteo Beccaria, Keith Dienes, Matthias Gaberdiel, Dileep Jatkar, Sunil Mukhi, Ashoke Sen, Bogdan Stefanski and G.M.T. Watts for discussions. We are grateful to Kris Thielemans for providing us with the Mathematica packages OPEdefs and OPEconf. D.K thanks IISER, Mohali and M.S thanks King's College, London, IISc, Bangalore, ICTS, Bangalore and City University, London for hospitality during different stages of this work. 
\end{acknowledgements}
\appendix

\section{Notation for ${SO}(2N)$}\label{Fso}

In this appendix, we state our notation for the $SO(2N)$ group.  
We will work in an orthonormal basis:~$\ve_1, \dots, \ve_N.$
In this basis the simple roots of $SO(2N)$ are
\begin{equation}
\begin{split}
\alpha_i &= \ve_i - \ve_{i+1}  \;\;\; \textrm{for}  \;\;\; 1 \leq i \leq N-1\,, \\
\alpha_N &= \ve_{N-1} + \ve_N\ .
\end{split}
\end{equation}
The fundamental weights are 
\begin{equation}
\begin{split}
\lambda_i &= \ve_1 + \ve_2 + \dots + \ve_{i}  \;\;\; \textrm{for}  \;\;\;  1\leq i \leq N-2 \,,\\
\lambda_{N-1} &= \tfrac{1}{2} \left( \ve_1 + \ve_2 + \dots + \ve_{N-1} - \ve_{N} \right)\,, \\
\lambda_{N} &= \tfrac{1}{2} \left( \ve_1 + \ve_2 + \dots + \ve_{N-1} + \ve_{N} \right) \,.
\end{split}
\end{equation}
The Weyl vector $\rho$ is given by
\begin{equation}\label{Weylv}
\rho = \sum_{i=1}^{N} \lambda_i =  \sum_{i = 1}^N (N -i) \ve_i.
\end{equation}

In terms of the fundamental weights, the weight $\Lambda$ of a highest weight representation (hwr) is:
\begin{equation}
\L = \sum_{p = 1}^N \L_p \lambda_p\ , 
\end{equation}
where $\L_p\geq 0$ are the Dynkin labels of $\L$. 
 In terms of the orthonormal basis $\L$ can be expanded as
 \begin{equation}\label{LamONB}
\L = \sum_{p = 1}^N l_i \ve_i\ .
\end{equation}
These expansion coefficients $l_i$ can be expressed in terms of the Dynkin labels:
\begin{equation}\label{lidef}
\begin{split}
l_i &= \sum_{p = i}^{N-2} \L_p + \tfrac{1}{2} \left( \L_{N-1} + \L_{N} \right)  \;\;\; \textrm{for}  \;\;\; 1\leq i \leq N-2 \\
l_{N-1} &= \tfrac{1}{2} \left( \L_{N-1} + \L_{N} \right) \ , \qquad l_{N} = \tfrac{1}{2} \left( \L_{N} - \L_{N-1} \right)\ .
\end{split}
\end{equation}
For a highest weight representation $\L$, the quadratic Casimir is
\begin{equation}\label{quadC}
C_N(\L) = \frac{1}{2} \left\langle \L, \L + 2 \rho \right\rangle 
= \frac{1}{2} \sum_{i = 1}^N l_i^2 + \sum_{i = 1}^N l_i \, (N-i) \ .
\end{equation}
Since we are working in the large $N$ limit, we need only work with representations for which the quadratic 
Casimir grows linearly with $N$. These representations satisfy
$\L_{N-1}=\L_N=0$.  From \equ{lidef},  we can see that these representations can be labelled by $l_i$, with $l_i\geq l_{i+1}$ for $i=1,\ldots, N-2$, 
and $l_{N-1}=l_{N}= 0$. Since the $l_i$ are non-negative ordered integers, we can interpret them as the number of boxes in the
$i^{\rm th}$ row of a Young tableaux. Let $c_j$ denote the number of boxes in the $j^{\rm th}$ column of such a Young tableaux.
Then the quadratic Casimir of a weight $\L$ corresponding to this Young diagram is
\begin{equation}\label{A.18}
C_N(\L) = \tfrac{1}{2} \left\langle \L, \L + 2 \rho \right\rangle 
= \tfrac{1}{2} \sum_{i = 1}^N l_i^2 + \sum_{i = 1}^N l_i \, (N-i)
= B (\L) \left(N - \tfrac{1}{2} \right) + \tfrac{1}{2} D(\L),
\end{equation}
where
\begin{equation}\label{A.17}
B(\L) = \sum_{i = 1}^{N} l_i \ , \qquad D(\L) = \sum_{i=1}^N l_i^2 - \sum_{i = 1}^N c_i^2 \ . 
\end{equation}

\subsection{Frobenius notation}
\begin{figure}[t!] 
\begin{center}
\includegraphics[scale=.45]{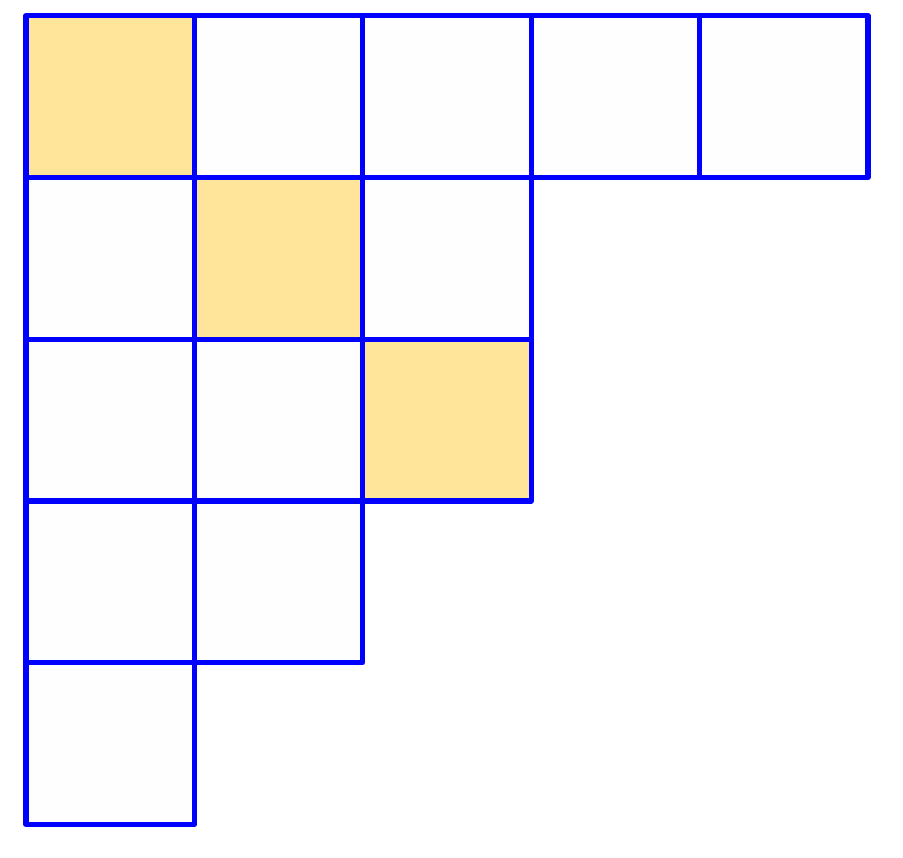}
\caption{\label{youngfig} Young diagram with Frobenius coordinates given by the matrix in \equ{fmatrix}. The shaded boxes represent the main diagonal.}
\end{center}
\end{figure}

In this paper we use the Frobenius notation for Young diagrams. In any Young diagram, let $d$ represent the number of boxes in the main diagonal. The Young diagram is, then, labelled by two sets of integers $a_i$ and $b_i$, where $i$ goes from $1$ to $d$. For the $i$th box on the diagonal, $a_i$ is the number of boxes to the right while $b_i$ is the number of boxes below. This is usually represented by a $2\times d$ array of integers, with the integers $a_i$ in the top row and the integers $b_i$ in the bottom row. For example, for the Young diagram in Figure~(\ref{youngfig}), the array is 
\begin{equation}\label{fmatrix}
\begin{pmatrix} 
4 & 1&0 \\
4 & 2 &0 
\end{pmatrix}
\end{equation}

For a representation $\L$, for which the $i$th row has $l_i$ boxes and the $i$th column has $c_i$ boxes, the corresponding Frobenius coordinates $a_i$ and $b_i$ are
\be
a_i= l_i - i,~~b_i = c_i-i~\,\,\textrm{where}~i=1~\textrm{to}~d\,.
\ee
The quadratic Casimir is given by \equ{A.18}, where $B(\L)$ is 
\be \label{BF}
  B(\L) = \sum_{i = 1}^{d} (a_i + b_i + 1) 
\ee 
and $D(\L)$ is 
\be \label{DF}
	D(\L) = \sum_{i=1}^d \(a_i + \tfrac{1}{2}\)^2 - \sum_{i = 1}^d \(b_i + \tfrac{1}{2}\)^2 \,.
\ee

\section{Branching function of the $D_N$ coset \label{gco}}
For completeness, we write down the branching function of the coset in \equ{sdncoset}, in the large $N$ limit. For details, the reader is referred to \cite{Gaberdiel:2011nt}.
For the diagonal coset $\frac{SO(2N)_{k}\otimes SO(2N)_{1}}{SO(2N)_{k+1}}$, let  $\Lambda_{+}$
and $\Lambda_{-}$ denote hwr of $SO(2N)_{k}$ and $SO(2N)_{k+1}$
respectively. Then the corresponding branching function $b_{(\Lambda_{+};\Lambda_{-})}$ is \cite{Bouwknegt:1992}
\be
b_{(\Lambda_{+};\Lambda_{-})}(q)= \frac{1}{\eta(q)^{N}}\sum_{w\in W} \sum_{ \substack{n_i \in \mathbb Z\\i= 1,\cdots, N}}\epsilon(w) q^{\frac{1}{2p(p+1)}\left|(p+1)(w(\Lambda_{+}+\rho)+p \sum_{i=1}^{N}n_{i}\alpha_{i}^{\vee})- p(\Lambda_{-}+\rho)\right|^{2}}\,,
\ee
where $\rho$ is the Weyl vector of $SO(2N)$, $\alpha_{i}^{\vee}$ are the co-roots, $W$ is the Weyl group and $p \equiv k+h$ where $h=2N-2$. In the limit of large $N$, we can neglect
the sum over the coroot lattice as the contribution of corresponding terms is exponentially suppressed. Therefore, we are left with 
\begin{equation}
b_{(\Lambda_{+};\Lambda_{-})}(q) \cong \frac{1}{\eta(q)^{N}}\sum_{w\in W}\epsilon(w)q^{\frac{1}{2p(p+1)}\left|(p+1)w(\Lambda_{+}+\rho)-p(\Lambda_{-}+\rho)\right|^{2}}\,.
\end{equation}
For the special case of $k=2N-2$ and in the large $N$ limit the term in the exponential can be written as:
\begin{gather}
C(\Lambda_{+})+C(\Lambda_{-})+ \tfrac{1}{4} B(\Lambda_{+}) - \tfrac{1}{4} B(\Lambda_{-}) +\left(1+\frac{1}{2p(p+1)}\right)\rho^{2}-\langle w(\Lambda_{+}+\rho),\Lambda_{-}+\rho\rangle \,,
\end{gather}
where $C(\Lambda)=\frac{1}{2}\langle\Lambda,\Lambda+2\rho\rangle$.
The branching function becomes
\begin{align}
b_{(\Lambda_{+};\Lambda_{-})}(q) & \cong  \frac{q^{\left(1+\frac{1}{2p(p+1)}\right)\rho^{2}}}{\eta(\tau)^{N}}  q^{C(\Lambda_{+})+C(\Lambda_{-})} q^{\frac{1}{4}B(\Lambda_{+})-\frac{1}{4}B(\Lambda_{-})}\sum_{w\in{W}}\epsilon(w)q^{-\langle w(\Lambda_{+}+\rho),\:\Lambda_{-}+\rho\rangle}\,.
\end{align}
This can be rearranged to
\begin{align}
\label{brandetail}
 b_{(\Lambda_{+};\Lambda_{-})}(q) = q^{\frac{1}{4}B(\Lambda_{+})-\frac{1}{4}B(\Lambda_{-})} & \sum_{w\in{W}}\epsilon(w)q^{-\langle w(\rho),\:\rho\rangle}  \frac{q^{\rho^{2}-c/24}}{  \prod_{n=1}^{\infty}(1-q^{n})^{N}} ~\times~ \nn \\  
 &\sum_{\Lambda} q^{ C(\Lambda)} N_{\Lambda_+ \Lambda_-}^{\Lambda} \frac{\sum_{w\in{W}}\epsilon(w)q^{-\langle w(\Lambda +\rho),\:\rho\rangle}}{\sum_{w\in{W}}\epsilon(w)q^{-\langle w(\rho),\:\rho\rangle}}\,,
\end{align}
where we have used $c=N-\frac{12}{p(p+1)}\rho^{2}$. The $N_{\Lambda_+ \Lambda_-}^{\Lambda}$ denote the Clebsch-Gordon coefficients for $SO(2N)$. By the Weyl denominator formula 
\begin{eqnarray}
	\sum_{w\in{W}}\epsilon(w)q^{-\langle w(\Lambda+\rho),\:\rho\rangle} & = & q^{-\rho^{2}-\langle\Lambda,\:\rho\rangle}\prod_{i=2}^{N}\prod_{j=1}^{i-1}(1-q^{l_{j}-l_{i}+i-j})(1-q^{l_{j}+l_{i}+2N-i-j})
\end{eqnarray}
where the $l_i$'s are length of the rows of the Young tableau corresponding to the weight $\Lambda$. In the large $N$ limit
\begin{equation}
\frac{\sum_{w\in{W}}\epsilon(w)q^{-\langle w(\Lambda+\rho),\:\rho\rangle}}{\sum_{w\in{W}}\epsilon(w)q^{-\langle w(\rho),\:\rho\rangle}} \cong \frac{q^{-\langle\Lambda,\:\rho\rangle}\prod_{i=2}^{N}\prod_{j=1}^{i-1}(1-q^{l_{j}-l_{i}+i-j})}{\prod_{i=2}^{N}\prod_{j=1}^{i-1}(1-q^{i-j})}\,.
\end{equation}
Also, 
\begin{eqnarray}
\prod_{n=1}^{\infty}(1-q^{n})^{-N}\sum_{w\in{W}}\epsilon(w)q^{-\langle w(\rho),\:\rho\rangle} & = &  q^{-\rho^{2}}\prod_{\substack{s=2\\s\text{ even}}}^{\infty}\prod_{n=s}^{\infty}\frac{1}{1-q^{n}}=q^{-\rho^{2}}M^{e}(q)\,.
\end{eqnarray}
Here, $M^e(q)$ is the modified MacMahon function defined in \equ{MMe}. In the large $N$ limit, we should include those $\L$ in the summation over $\L$ in \equ{brandetail} for which $B(\L)=B(\L_+) +B(\L_-)$. The branching function then becomes
\begin{equation}
b_{(\Lambda_{+};\Lambda_{-})}(q)=q^{-\frac{c}{24}}\, M^e (q)  \, 
P^+_{Y^T(\L_+)} (q) P^-_{Y^T(\L_-)} (q)\,,
\end{equation}
where $P^{\pm}_{Y^T(\L_{\pm})}(q)$ are the modified Schur functions defined in \equ{Schur} with $\l=1/2$.
\section{Branching rules and series expansions \label{seriesEx}}
In this appendix, we construct the $AE$ and $EA$ partition functions for the Series~I and Series~II coset models for the $N=3$ case by determining the fields that should appear in the partition functions.  We also write down the $q$-series expansions for the associated branching functions of these fields and clarify the relation between the WZW weights of the fields that appear in the $AE$ and $EA$ partitions.
The $SU(3)_5$ and $SU(4)_2$ WZW groups are both embedded in $SU(6)_1$. The $SU(6)_1$ WZW model has six primary fields. The branching rules for the weights of the primary fields of the $SU(6)_1$ model, denoted by $\lambda$ in \sct{prop}, into the weights of the primary fields of the $SU(3)_5$ model, denoted by $\delta$ are as follows:
\begin{align}
\label{alphabranch}
	\l: (1,1,1,1,1) &\longrightarrow (1,1)+(3,3) \,,\nn\\
	\l: (2,1,1,1,1) &\longrightarrow (3,1)+(3,4) \,,\nn\\
	\l: (1,2,1,1,1) &\longrightarrow (1,6)+(3,2) \,,\nn\\
	\l: (1,1,2,1,1) &\longrightarrow (1,4)+(4,1) \,,\nn\\
	\l: (1,1,1,2,1) &\longrightarrow (6,1)+(2,3) \,,\nn\\
	\l: (1,1,1,1,2) &\longrightarrow (1,3)+(4,3)\,.
\end{align}
To get the weights of the primary fields that constitute the $AE$ partition function for the Series~I coset, the weights appearing on the right hand side in the above equations need to be paired with the weights of $SU(3)_4$, which are denoted by $\alpha$. However, not all pairs of weights $(\alpha, \delta)$ will appear in the partition function --- only pairs that are distinct after field identification.  We can choose the pairs in such a way that we keep all the weights $\delta$ but restrict the weights $\alpha$ when using field identification. Then, the number of distinct weights of $SU(3)_4$ appearing in the partition function is $5$ and one choice for these weights is
\be
\label{alphawts1}
	\alpha \in \left  \{(1,1),(1,2),(1,3),(1,4),(2,2)\right\}\,.
\ee
In Table~(\ref{apptab1}) we list the weights $(\a,\delta)$ of the fields that arise in the $AE$ partition for the $N=3$ Series~I coset along with their $q$-series expansions.  Note that there is a associated value of $\l$ with each row that can be read off from \equ{alphabranch} by matching the $\delta$ value for a particular row with the $\delta$ value appearing in the R.H.S of \equ{alphabranch}. To match with the partition function in \equ{eq:AE5}, we have included the conformal dimension of the fields: the fields listed in Table~(\ref{apptab1}) are in one-to-one correspondence with those appearing in \equ{eq:AE5}. For the cases where fields with the same conformal dimension have differing $q$-series expansion, we have also included the $W_3$ charge.

The branching rules for the weights of the primary fields of the $SU(6)_1$ model, now denoted by $\lambda'$, into the weights of the primary fields of the $SU(4)_2$ model, denoted by $\beta$ are as follows:
\begin{align}
\label{lambdabranch2}
	\l': (1,1,1,1,1) &\longrightarrow (1,1,1)+(1,3,1) \,,\nn\\
	\l': (2,1,1,1,1) &\longrightarrow (1,2,1) \,,\nn\\
	\l': (1,2,1,1,1) &\longrightarrow (2,1,2) \,,\nn\\
	\l': (1,1,2,1,1) &\longrightarrow (3,1,1)+(1,1,3) \,,\nn\\
	\l': (1,1,1,2,1) &\longrightarrow (2,1,2) \,,\nn\\
	\l': (1,1,1,1,2) &\longrightarrow (1,2,1)\,.
\end{align}
To get the primary fields that constitute the $EA$ partition function for the Series~II coset, we pair the weights appearing on the right hand side of the above equation with the weights of $SU(4)_3$,  denoted by $\alpha'$.  As for the previous case, we can again choose the pairs in such a way that the weight $\beta$ is unrestricted but the weight $\alpha'$ is restricted by field identification. This constraint is automatically satisfied if we choose the weights of $SU(4)_3$ that are related to the $SU(3)_4$ weights in \equ{alphawts1} by the level-rank duality map: that is transpose rows of the Young tableau for $\a$ into columns. The weights $\a'$ are then
\be
\label{alphawts2}
	\alpha' \in \{(1,1,1),(3,1,1),(1,3,1),(1,1,3),(2,2,1)\}\,.
\ee
In Table~(\ref{apptab2}) we list the weights $(\b,\a')$ of the fields that arise in the $AE$ partition for Series~II along with their $q$-series expansions. The associated $\l'$ value can be read off from \equ{lambdabranch2}. These weights are in one-to-one correspondence with those appearing in \equ{eq:EA5}. Because some $SU(4)_2$ weights appear twice on the R.H.S of the equations in (\ref{lambdabranch2}), Table~(\ref{apptab2}) has some degeneracies.

As noted below \equ{branchEquiv}, there is a one-to-one map from the set of weights $(\l,\a)$ to the set of weights $(\l',\a')$. The corresponding branching function identities for the $N=3$ case are given in \equ{n3Identities}. Note that, in general, $\l$ and $\l'$ are not equal. We clarify this by giving some examples. We can read the map for the identity $b_{\frac{1}{7}}^{\I\I}+b_{\frac{8}{7}}^{\I\I}~=~b_{\frac{1}{7}}^{\I}+b_{\frac{36}{7}}^{\I}$ from Tables~(\ref{apptab1}) and (\ref{apptab2}) and Eqs.~(\ref{alphabranch}) and (\ref{lambdabranch2}). It is given by:
\be
	(\l,\a): \{(1,1,1,2,1),(1,3)\} \longrightarrow (\l',\a'): \{(1,1,1,1,1),(1,3,1)\}\,.
\ee
The identity $b_{\frac{3}{28}}^{\I\I}+b_{\frac{59}{28}}^{\I\I}~=~b_{\frac{3}{28}}^{\I}+b_{\frac{87}{28}}^{\I}$ is two distinct identities in terms of the WZW labels of the primary fields and corresponds to the following two maps:
\begin{align}
	(\l,\a): \{(1,1,1,1,2),(1,2)\} &\longrightarrow (\l',\a'): \{(1,1,2,1,1),(3,1,1)\}\,,\nn\\
	(\l,\a): \{(1,1,2,1,1),(1,4)\} &\longrightarrow (\l',\a'): \{(1,1,2,1,1),(1,1,3)\}\,.
\end{align}
Similarly, the identity $b_{\frac{5}{84}}^{\I\I}~=~b_{\frac{5}{84}}^{\I}+b_{\frac{173}{84}}^{\I}$ is also two distinct identities corresponding, for example, to the maps:
\begin{align}
	(\l,\a): \{(1,1,1,1,2),(1,3)\} &\longrightarrow (\l',\a'): \{(1,1,1,1,2),(1,3,1)\}\,,\nn\\
	(\l,\a): \{(1,1,2,1,1),(1,3)\} &\longrightarrow (\l',\a'): \{(2,1,1,1,1),(1,3,1)\}\,.
\end{align}
As can be seen above, the indices $\l$ and $\l'$ are not always equal. This is because as stated in \sct{prop}, the identities depend on the conformal dimensions of the fields involved. The conformal dimension of a coset primary in terms of the generic weights $(\L_+,\L_-) $ of the constituent WZW models is given by
\be
\label{cdimension}
	h(\L_+,\L_-) = \frac{1}{2 \, r\, (r+1)}\left( \big | (r+1) (\L_+  + \hat{\rho}) -  r (\L_- + \hat{\rho}) \big |^2 - \hat{\rho}^2 \right)\,,
\ee
where $r=N+k$ and $\hat{\rho}$ and is the Weyl vector for $SU(N)$. 
\begin{center}
\setlength\extrarowheight{2pt}
 \begin{ruledtabular}
\begin{longtable}{|c|c|l|}
\caption{\label{apptab1}\large{ Branching functions for the primary fields of the $AE$ partition function of the coset} $\tfrac{SU(3)_4\otimes SU(3)_1}{SU(3)_5}$}  \\
\hline
\textbf{Weight $(\a, \delta)$} & \textbf{Conformal weight } & \textbf{Branching function $b_{\a\delta}(q)$} \\
\hline
\endfirsthead
\hline
\textbf{Weight $(\a,\delta)$} & \textbf{Conformal weight } & \textbf{Branching function $b_{\a \delta}(q)$} \\
\hline 
\endhead 
\multicolumn{3}{r}{\textit{\vspace{0.2cm}Continued on next page ...}} \\
\endfoot
\endlastfoot
$(1,1),(1,1)$ & $0$ & $q^{-\frac{11}{168}}(1+q^{2}+2q^{3}+3q^{4}+4q^{5}+\dots)$\tabularnewline
\hline 
$(1,1),(3,3)$ & $3$ & $q^{\frac{493}{168}}(1+2q+5q^{2}+8q^{3}+16q^{4}+26q^{5}+\dots)$\tabularnewline
\hline 
$(1,2),(1,3)$ & \multirow{2}{*}{$\frac{3}{28}$} & \multirow{2}{*}{$q^{\frac{1}{24}}(1+q+3q^{2}+5q^{3}+10q^{4}+16q^{5}+\dots)$}\tabularnewline
\cline{1-1} 
$(1,4),(1,4)$ &  & \tabularnewline
\hline 
$(1,2),(4,3)$ & \multirow{2}{*}{$\frac{87}{28}$} & \multirow{2}{*}{$q^{\frac{73}{24}}(1+2q+5q^{2}+10q^{3}+18q^{4}+32q^{5}+\dots)$}\tabularnewline
\cline{1-1} 
$(1,4),(4,1)$ &  & \tabularnewline
\hline 
$(1,3),(1,3)$ & \multirow{2}{*}{$\frac{5}{84}$} & \multirow{2}{*}{$q^{-\frac{1}{168}}(1+q+3q^{2}+5q^{3}+10q^{4}+16q^{5}+\dots)$}\tabularnewline
\cline{1-1} 
$(1,3),(1,4)$ &  & \tabularnewline
\hline 
$(1,3),(4,1)$ & \multirow{2}{*}{$\frac{173}{84}$} & \multirow{2}{*}{$q^{\frac{335}{168}}(1+2q+5q^{2}+9q^{3}+17q^{4}+29q^{5}+\dots)$}\tabularnewline
\cline{1-1} 
$(1,3),(4,3)$ &  & \tabularnewline
\hline 
$(2,2),(2,3)$ & \multirow{2}{*}{$\frac{2}{21}$} & \multirow{2}{*}{$q^{\frac{5}{168}}(1+2q+5q^{2}+10q^{3}+19q^{4}+34q^{5}+\dots)$}\tabularnewline
\cline{1-1} 
$(2,2),(3,2)$ &  & \tabularnewline
\hline 
$(2,2),(1,6)$ & \multirow{2}{*}{$\frac{65}{21}$} & \multirow{2}{*}{$q^{\frac{509}{168}}(1+2q+4q^{2}+7q^{3}+13q^{4}+21q^{5}+\dots)$}\tabularnewline
\cline{1-1} 
$(2,2),(6,1)$ &  & \tabularnewline
\hline 
$(1,3),(2,3)$ & $\frac{1}{7}$ & $q^{\frac{13}{168}}(1+2q+4q^{2}+8q^{3}+15q^{4}+26q^{5}+\dots)$\tabularnewline
\hline 
$(1,3),(6,1)$ & $\frac{36}{7}$ & $q^{\frac{853}{168}}(1+2q+5q^{2}+8q^{3}+16q^{4}+26q^{5}+\dots)$\tabularnewline
\hline 
$(2,2),(1,3)$ & \multirow{2}{*}{$\frac{29}{84}$} & \multirow{2}{*}{$q^{\frac{47}{168}}(1+2q+4q^{2}+8q^{3}+15q^{4}+26q^{5}+\dots)$}\tabularnewline
\cline{1-1} 
$(2,2),(3,1)$ &  & \tabularnewline
\hline 
$(2,2),(3,4)$ & \multirow{2}{*}{$\frac{113}{84}$} & \multirow{2}{*}{$q^{\frac{215}{168}}(1+2q+5q^{2}+9q^{3}+18q^{4}+31q^{5}+\dots)$}\tabularnewline
\cline{1-1} 
$(2,2),(4,3)$ &  & \tabularnewline
\hline 
$(2,2),(3,3)$ & $\frac{3}{7}$ & $q^{\frac{61}{168}}(1+2q+5q^{2}+10q^{3}+20q^{4}+36q^{5}+\dots)$\tabularnewline
\hline 
$(2,2),(1,1)$ & $\frac{10}{7}$ & $q^{\frac{229}{168}}(1+2q+3q^{2}+6q^{3}+10q^{4}+16q^{5}+\dots)$\tabularnewline
\hline 
$(1,2),(1,1)$ & \multirow{2}{*}{$\frac{11}{21},+$} & \multirow{2}{*}{$q^{\frac{11}{24}}(1+q+2q^{2}+3q^{3}+6q^{4}+9q^{5}+\dots)$}\tabularnewline
\cline{1-1} 
$(1,4),(1,6)$ &  & \tabularnewline
\hline
$(1,2),(2,3)$ &  \multirow{2}{*}{$\frac{11}{21},-$} & \multirow{2}{*}{$q^{\frac{11}{24}}(1+2q+4q^{2}+8q^{3}+15q^{4}+26q^{5}+\dots)$}\tabularnewline
\cline{1-1} 
$(1,4),(2,3)$ &  & \tabularnewline
\hline 
$(1,2),(3,3)$ & \multirow{2}{*}{$\frac{32}{21}$} & \multirow{2}{*}{$q^{\frac{35}{24}}(1+2q+5q^{2}+9q^{3}+18q^{4}+31q^{5}+\dots)$}\tabularnewline
\cline{1-1} 
$(1,4),(3,2)$ &  & \tabularnewline
\hline 
$(1,2),(6,1)$ & \multirow{2}{*}{$\frac{116}{21}$} & \multirow{2}{*}{$q^{\frac{131}{24}}(1+2q+4q^{2}+8q^{3}+14q^{4}+24q^{5}+\dots)$}\tabularnewline
\cline{1-1} 
$(1,4),(6,1)$ &  & \tabularnewline
\hline 
$(1,2),(1,4)$ & \multirow{2}{*}{$\frac{65}{84},+$} & \multirow{2}{*}{$q^{\frac{17}{24}}(1+q+3q^{2}+5q^{3}+10q^{4}+16q^{5}+\dots)$}\tabularnewline
\cline{1-1} 
$(1,4),(1,3)$ &  & \tabularnewline
\hline
$(1,2),(3,1)$ & \multirow{2}{*}{$\frac{65}{84},-$}  & \multirow{2}{*}{$q^{\frac{17}{24}}(1+2q+4q^{2}+7q^{3}+13q^{4}+21q^{5}+\dots)$}\tabularnewline
\cline{1-1} 
$(1,4),(3,4)$ &  & \tabularnewline
\hline 
$(1,2),(4,1)$ & \multirow{2}{*}{$\frac{149}{84}$} & \multirow{2}{*}{$q^{\frac{41}{24}}(1+2q+4q^{2}+8q^{3}+14q^{4}+24q^{5}+\dots)$}\tabularnewline
\cline{1-1} 
$(1,4),(4,3)$ &  & \tabularnewline
\hline 
$(1,2),(3,4)$ & \multirow{2}{*}{$\frac{233}{84}$} & \multirow{2}{*}{$q^{\frac{65}{24}}(1+2q+5q^{2}+9q^{3}+17q^{4}+29q^{5}+\dots)$}\tabularnewline
\cline{1-1} 
$(1,4),(3,1)$ &  & \tabularnewline
\hline 
$(1,3),(3,2)$ & \multirow{2}{*}{$\frac{17}{21}$} & \multirow{2}{*}{$q^{\frac{125}{168}}(1+2q+5q^{2}+9q^{3}+18q^{4}+31q^{5}+\dots)$}\tabularnewline
\cline{1-1} 
$(1,3),(3,3)$ &  & \tabularnewline
\hline 
$(1,3),(1,1)$ & \multirow{2}{*}{$\frac{38}{21}$} & \multirow{2}{*}{$q^{\frac{293}{168}}(1+q+3q^{2}+4q^{3}+8q^{4}+12q^{5}+\dots)$}\tabularnewline
\cline{1-1} 
$(1,3),(1,6)$ &  & \tabularnewline
\hline 
$(1,2),(3,2)$ & \multirow{2}{*}{$\frac{6}{7}$} & \multirow{2}{*}{$q^{\frac{19}{24}}(1+2q+5q^{2}+9q^{3}+17q^{4}+29q^{5}+\dots)$}\tabularnewline
\cline{1-1} 
$(1,4),(3,3)$ &  & \tabularnewline
\hline 
$(1,2),(1,6)$ & \multirow{2}{*}{$\frac{27}{7}$} & \multirow{2}{*}{$q^{\frac{91}{24}}(1+q+3q^{2}+5q^{3}+9q^{4}+14q^{5}+\dots)$}\tabularnewline
\cline{1-1} 
$(1,4),(1,1)$ &  & \tabularnewline
\hline 
$(1,1),(1,3)$ & \multirow{2}{*}{$\frac{11}{12}$} & \multirow{2}{*}{$q^{\frac{143}{168}}(1+q+3q^{2}+4q^{3}+8q^{4}+12q^{5}+\dots)$}\tabularnewline
\cline{1-1} 
$(1,1),(3,1)$ &  & \tabularnewline
\hline 
$(1,1),(3,4)$ & \multirow{2}{*}{$\frac{59}{12}$} & \multirow{2}{*}{$q^{\frac{815}{168}}(1+2q+5q^{2}+9q^{3}+17q^{4}+28q^{5}+\dots)$}\tabularnewline
\cline{1-1} 
$(1,1),(4,3)$ &  & \tabularnewline
\hline 
$(1,1),(2,3)$ & \multirow{2}{*}{$\frac{5}{3}$} & \multirow{2}{*}{$q^{\frac{269}{168}}(1+2q+4q^{2}+7q^{3}+13q^{4}+21q^{5}+\dots)$}\tabularnewline
\cline{1-1} 
$(1,1),(3,2)$ &  & \tabularnewline
\hline 
$(1,1),(1,6)$ & \multirow{2}{*}{$\frac{20}{3}$} & \multirow{2}{*}{$q^{\frac{1109}{168}}(1+q+3q^{2}+5q^{3}+10q^{4}+15q^{5}+\dots)$}\tabularnewline
\cline{1-1} 
$(1,1),(6,1)$ &  & \tabularnewline
\hline 
$(2,2),(1,4)$ & \multirow{2}{*}{$\frac{19}{28}$} & \multirow{2}{*}{$q^{\frac{103}{168}}(1+2q+4q^{2}+8q^{3}+15q^{4}+26q^{5}+\dots)$}\tabularnewline
\cline{1-1} 
$(2,2),(4,1)$ &  & \tabularnewline
\hline 
$(1,3),(3,1)$ & \multirow{2}{*}{$\frac{39}{28}$} & \multirow{2}{*}{$q^{\frac{223}{168}}(1+2q+5q^{2}+8q^{3}+16q^{4}+26q^{5}+\dots)$}\tabularnewline
\cline{1-1} 
$(1,3),(3,4)$ &  & \tabularnewline
\hline 
$(1,1),(1,4)$ & \multirow{2}{*}{$\frac{9}{4}$} & \multirow{2}{*}{$q^{\frac{367}{168}}(1+q+3q^{2}+5q^{3}+9q^{4}+14q^{5}+\dots)$}\tabularnewline
\cline{1-1} 
$(1,1),(4,1)$ &  & \tabularnewline 
\hline 
\end{longtable}
\end{ruledtabular}
\end{center}

\begin{center}
\begin{longtable}{|c|c|l|}
\caption{ \label{apptab2} \large{Branching functions of primary fields of the $EA$ partition function for the coset $\tfrac{SU(4)_2\otimes SU(4)_1}{SU(4)_3}$}}\\
\hline
\textbf{Weight $(\beta,\alpha')$} & \textbf{Conformal weight } & \textbf{Branching function $b_{\beta \alpha'}(q)$} \\
\hline
\endfirsthead
\hline
\textbf{Weight $(\beta,\alpha')$} & \textbf{Conformal weight } & \textbf{Branching function $b_{\beta \alpha'}(q)$} \\
\hline 
\endhead
\multicolumn{3}{r}{\textit{Continued on next page ...}} \\
\endfoot
\endlastfoot
$(1,1,1),\:(1,1,1)$ & $0$ & $q^{-\frac{11}{168}}(1+q^{2}+2q^{3}+4q^{4}+5q^{5}+\dots)$\tabularnewline
\hline 
$(1,3,1),\:(1,1,1)$ & $3$ & $q^{\frac{493}{168}}(1+q+4q^{2}+6q^{3}+12q^{4}+18q^{5}+\dots)$\tabularnewline
\hline 
$(3,1,1),\:(3,1,1)$ & \multirow{2}{*}{$\frac{3}{28}$} & \multirow{2}{*}{$q^{\frac{1}{24}}(1+q+2q^{2}+4q^{3}+7q^{4}+12q^{5}+\dots)$}\tabularnewline
\cline{1-1} 
$(1,1,3),\:(1,1,3)$ &  & \tabularnewline
\hline 
$(3,1,1),\:(1,1,3)$ & \multirow{2}{*}{$\frac{59}{28}$} & \multirow{2}{*}{$q^{\frac{49}{24}}(1+2q+5q^{2}+9q^{3}+17q^{4}+28q^{5}+\dots)$}\tabularnewline
\cline{1-1} 
$(1,1,3),\:(3,1,1)$ &  & \tabularnewline
\hline 
$(1,2,1),\:(1,3,1)$ & \multirow{2}{*}{$\frac{5}{84}$} & \multirow{2}{*}{$q^{-\frac{1}{168}}(1+q+4q^{2}+7q^{3}+15q^{4}+25q^{5}+\dots)$}\tabularnewline
\cline{1-1} 
$(1,2,1),\:(1,3,1)$ &  & \tabularnewline
\hline 
$(1,2,1),\:(2,2,1)$ & \multirow{2}{*}{$\frac{2}{21}$ }&  \multirow{2}{*}{$q^{\frac{5}{168}}(1+2q+5q^{2}+11q^{3}+21q^{4}+38q^{5}+\dots)$}\tabularnewline
\cline{1-1} 
$(1,2,1),\:(2,2,1)$ &  & \tabularnewline
\hline 
$(1,3,1),\:(1,3,1)$ &  $\frac{1}{7}$ & $q^{\frac{13}{168}}(1+q+3q^{2}+4q^{3}+9q^{4}+14q^{5}+\dots)$\tabularnewline
\hline 
$(1,1,1),\:(1,3,1)$ & $\frac{8}{7}$ & $q^{\frac{181}{168}}(1+q+4q^{2}+6q^{3}+13q^{4}+21q^{5}+\dots)$\tabularnewline
\hline 
$(2,1,2),\:(2,2,1)$ & \multirow{2}{*}{$\frac{29}{84}$} &  \multirow{2}{*}{$q^{\frac{47}{168}}(1+3q+6q^{2}+13q^{3}+24q^{4}+44q^{5}+\dots)$}\tabularnewline
\cline{1-1} 
$(2,1,2),\:(2,2,1)$ &  & \tabularnewline
\hline 
$(3,1,1),\:(2,2,1)$ & $\frac{3}{7}$ & $q^{\frac{61}{168}}(1+2q+4q^{2}+7q^{3}+14q^{4}+24q^{5}+\dots)$\tabularnewline
\hline 
$(1,1,3),\:(2,2,1)$ & $\frac{10}{7}$ & $q^{\frac{229}{168}}(1+3q+6q^{2}+12q^{3}+22q^{4}+37q^{5}+\dots)$\tabularnewline
\hline 
$(2,1,2),\:(3,1,1)$ & \multirow{4}{*}{$\frac{11}{21}$} & \multirow{4}{*}{$q^{\frac{11}{24}}(1+2q+4q^{2}+8q^{3}+15q^{4}+27q^{5}+\dots)$}\tabularnewline
\cline{1-1} 
$(2,1,2),\:(3,1,1)$ &  & \tabularnewline
\cline{1-1} 
$(2,1,2),\:(1,1,3)$ &  & \tabularnewline
\cline{1-1} 
$(2,1,2),\:(1,1,3)$ &  & \tabularnewline
\hline 
$(1,3,1),\:(2,2,1)$ & \multirow{2}{*}{$\frac{19}{28}$} & \multirow{2}{*}{$q^{\frac{103}{168}}(1+2q+4q^{2}+8q^{3}+15q^{4}+26q^{5}+\dots)$}\tabularnewline
\cline{1-1} 
$(1,1,1),\:(2,2,1)$ &  & \tabularnewline
\hline 
$(1,2,1),\:(1,1,3)$ & \multirow{4}{*}{$\frac{65}{84}$} & \multirow{4}{*}{$q^{\frac{17}{24}}(1+2q+5q^{2}+9q^{3}+18q^{4}+30q^{5}+\dots)$}\tabularnewline
\cline{1-1} 
$(1,2,1),\:(1,1,3)$ &  & \tabularnewline
\cline{1-1} 
$(1,2,1),\:(3,1,1)$ &  & \tabularnewline
\cline{1-1} 
$(1,2,1),\:(3,1,1)$ &  & \tabularnewline
\hline 
$(2,1,2),\:(1,3,1)$ & \multirow{2}{*}{$\frac{17}{21}$ }&\multirow{2}{*}{ $q^{\frac{125}{168}}(1+3q+6q^{2}+12q^{3}+22q^{4}+39q^{5}+\dots)$}\tabularnewline
\cline{1-1} 
$(2,1,2),\:(1,3,1)$ &  & \tabularnewline
\hline 
$(1,1,1),\:(1,1,3)$ & \multirow{2}{*}{$\frac{6}{7}$} & \multirow{2}{*}{$q^{\frac{19}{24}}(1+q+3q^{2}+5q^{3}+10q^{4}+16q^{5}+\dots)$}\tabularnewline
\cline{1-1} 
$(1,1,1),\:(3,1,1)$ &  & \tabularnewline
\hline 
$(1,3,1),\:(3,1,1)$ & \multirow{2}{*}{$\frac{13}{7}$} & \multirow{2}{*}{$q^{\frac{43}{24}}(1+2q+5q^{2}+8q^{3}+16q^{4}+26q^{5}+\dots)$}\tabularnewline
\cline{1-1} 
$(1,3,1),\:(1,1,3)$ &  & \tabularnewline
\hline 
$(1,2,1),\:(1,1,1)$ & \multirow{2}{*}{$\frac{11}{12}$} & \multirow{2}{*}{$q^{\frac{143}{168}}(1+q+3q^{2}+4q^{3}+9q^{4}+14q^{5}+\dots)$}\tabularnewline
\cline{1-1} 
$(1,2,1),\:(1,1,1)$ &  & \tabularnewline
\hline 
$(1,1,3),\:(1,3,1)$ & \multirow{2}{*}{$\frac{39}{28}$} & \multirow{2}{*}{$q^{\frac{223}{168}}(1+2q+5q^{2}+8q^{3}+16q^{4}+26q^{5}+\dots)$}\tabularnewline
\cline{1-1} 
$(3,1,1),\:(1,3,1)$ &  & \tabularnewline
\hline 
$(2,1,2),\:(1,1,1)$ &\multirow{2}{*}{ $\frac{5}{3}$ }& \multirow{2}{*}{$q^{\frac{269}{168}}(1+2q+4q^{2}+7q^{3}+13q^{4}+22q^{5}+\dots)$}\tabularnewline
\cline{1-1} 
$(2,1,2),\:(1,1,1)$ &  & \tabularnewline
\hline 
$(3,1,1),\:(1,1,1)$ & \multirow{2}{*}{$\frac{9}{4}$} & \multirow{2}{*}{$q^{\frac{367}{168}}(1+q+3q^{2}+5q^{3}+9q^{4}+14q^{5}+\dots)$}\tabularnewline
\cline{1-1} 
$(1,1,3),\:(1,1,1)$ &  & \tabularnewline
\hline 
\end{longtable}
\end{center}
	

\end{document}